%% file: main.tex
\pdfoutput=1
\RequirePackage{etoolbox}

\providebool{fullversion}
\providebool{arxivversion}

\booltrue{arxivversion}

\ifbool{arxivversion}{%
\documentclass[sigconf,screen,nonacm]{acmart}\settopmatter{printfolios=true,printccs=false,printacmref=false}
}{%
\documentclass[sigconf,anonymous,review,screen,nonacm]{acmart}\settopmatter{printfolios=true,printccs=false,printacmref=false}
}

\usepackage{preamble}
\usepackage{iris}
\usepackage{prob}
\usepackage{examples}

\usepackage{multicol}

\setcopyright{acmlicensed}
\copyrightyear{2025}
\acmYear{2025}
\acmDOI{XXXXXXX.XXXXXXX}

\ifbool{fullversion}{
  \newcommand{\appref}[1]{\cref{#1}}
  \newcommand{\Appref}[1]{\Cref{#1}}
}{
  \newcommand{\appref}[1]{the Appendix}
  \newcommand{\Appref}[1]{The Appendix}
}

\title[Logical Relations for Formally Verified Authenticated Data Structures]{Logical Relations for Formally Verified\\ Authenticated Data Structures}

\author{Simon Oddershede Gregersen}
\orcid{0000-0001-6045-5232}
\affiliation{
  \institution{New York University}
  \city{New York}
  \state{NY}
  \country{USA}
}
\email{s.gregersen@nyu.edu}

\author{Chaitanya Agarwal}
\orcid{}
\affiliation{
  \institution{New York University}
  \city{New York}
  \state{NY}
  \country{USA}
}
\email{ca2719@nyu.edu}

\author{Joseph Tassarotti}
\orcid{0000-0001-5692-3347}
\affiliation{
  \institution{New York University}
  \city{New York}
  \state{NY}
  \country{USA}
}
\email{jt4767@nyu.edu}

\begin{document}

\begin{abstract}
  Authenticated data structures allow untrusted third parties to carry out operations which produce proofs that can be used to verify an operation's output.
  Such data structures are challenging to develop and implement correctly.
  This paper gives a formal proof of security and correctness for a library that generates authenticated versions of data structures automatically.
  The proof is based on a new relational separation logic for reasoning about programs that use collision-resistant cryptographic hash functions.
  This logic provides a basis for constructing two semantic models of a type system, which are used to justify how the library makes use of type abstraction to enforce security and correctness.
  Using these models, we also prove the correctness of several optimizations to the library and then show how optimized, hand-written implementations of authenticated data structures can be soundly linked with automatically generated code.
  All of the results in this paper have been mechanized in the Coq proof assistant using the Iris framework.
\end{abstract}
\maketitle

\input{introduction}
\input{background}
\input{logic}
\input{logrel}
\input{optimizations}
\input{semantic-proofs}
\input{work}
\input{conclusion}

\begin{acks}
  This work was supported in part by the \grantsponsor{NSF}{National Science Foundation}{}, grant no.~\grantnum{NSF}{2338317} and the \grantsponsor{Carlsberg Foundation}{Carlsberg Foundation}{}, grant no.~\grantnum{Carlsberg Foundation}{CF23-0791}.
\end{acks}

\bibliography{refs}

\appendix

\input{appendix.tex}

\end{document}

%% file: introduction.tex
\section{Introduction}
\label{sec:introduction}

An \emph{authenticated data structure}~(ADS)~\citep{tamassia-ads} is a type of data structure in which operations produce cryptographic proofs that can be used to check that the results of the operation are valid.
ADSs can be used in scenarios where computations are outsourced to untrusted third parties, where the results of the computation can be verified by checking the proofs that are produced.
For example, Merkle trees~\citep{merkle-tree} are a binary tree data structure in which internal nodes of the tree contain hashes of their children.
A retrieve operation on a Merkle tree produces a proof consisting of a list of hashes of nodes.
A verifier that only knows the hash of the root node of the tree is able to check the result of the retrieve by re-computing what the root hash would be from the proof and comparing the result to the actual root hash.

Because ADSs are used for security-critical applications, it is essential that they are correctly implemented.
In particular, adversaries must not be able to construct false proofs that trick a verifier into accepting an invalid result.
However, correctly developing and implementing ADSs is challenging.
Moreover, it is difficult to test ADS implementations to ensure that adversarially-constructed invalid proofs are rejected.

To ease the task of developing ADSs, \citet{miller-ads} develop \lambdaAuth, an extension to the OCaml programming language for implementing ADSs that use \emph{cryptographic hash functions} to construct proofs.
The $\lambdaAuth$ language extends OCaml with a new class of \emph{authenticated types}, which describe data that comes with an accompanying cryptographic proof.
Using $\lambdaAuth$, a developer writes a standard implementation of a data structure, and then the compiler automatically generates code for an authenticated version.
Specifically, the $\lambdaAuth$ compiler converts a program into two different executables, called the \emph{prover} and \emph{verifier} executable.
The prover performs operations and generates cryptographic proofs corresponding to those operations.
Meanwhile, the verifier takes a proof as input and checks whether it is valid or not.
Experimental evaluations show that the $\lambdaAuth$ compiler can generate authenticated data structures with the same asymptotic running time as manually-written versions.

To justify the correctness of $\lambdaAuth$'s approach to generating authenticated data structures, \citet{miller-ads} present a core calculus language for a subset of $\lambdaAuth$ and prove security and correctness properties of programs written in this language.
This core calculus has separate formal operational semantics describing how the prover and verifier versions of a program should run.
In these different semantics, primitive operations in the language generate or produce proofs.
In addition, there is a third operational semantics for \emph{ideal} execution, in which there are no proof objects and programs execute like a standard, simplified OCaml-like program.
This ideal execution behavior is not intended to be run directly, but instead is used to specify how the prover and verifier should behave.
In particular, all well-typed programs in this language satisfy two important properties, called security and correctness, which can be informally summarized as:
\begin{itemize}
\item \textbf{Security}: If the verifier version of a program accepts a proof $p$ and returns a value $v$, then either the ideal execution of the program would have also returned $v$, or a hash collision must have occurred during the execution of the verifier.
\item \textbf{Correctness}: If the prover version generates a proof $p$ and returns a value $v$, then the ideal version would also return $v$, and the verifier will accept $p$ and return $v$ as well.
\end{itemize}
In particular, if the cryptographic hash function used in the authenticated data structure operations has (strong) collision resistance, then this security property ensures that it is hard for an attacker to construct a malicious proof that deceives a verifier.

However, there are three important limitations of the $\lambdaAuth$ approach.
First, maintaining a custom compiler frontend imposes development burden.
Second, to achieve efficient performance, the $\lambdaAuth$ compiler implements several optimizations related to how proof objects are stored and checked, but these optimizations are not covered by the security and correctness theorems, because the core calculus only models a simple, unoptimized version of proof object operations.
Third, even with these compiler optimizations, the data structures generated by $\lambdaAuth$ are not always as efficient as hand-written versions.
For example, \citeauthor{miller-ads} report that the $\lambdaAuth$-generated verifier for a binary tree structure takes twice as long to check proofs as a hand-written C version.

In subsequent work, \citet{atkey-ads} addressed the first of these limitations by showing that $\lambdaAuth$'s authenticated types could be directly encoded in OCaml's existing type system, without needing to extend the language or use a custom compiler frontend.
With this approach, a programmer uses OCaml's module system to write an implementation of a data structure that is \emph{parameterized} by functions that handle proof generation and checking.
Then, by linking the data structure code with three different libraries implementing these functions, one obtains the prover, verifier, and ideal versions of an authenticated data structure.
Although this approach eliminates the need for a custom compiler, it raises its own challenges.
Whereas \citeauthor{miller-ads} were able to prove security and correctness of $\lambdaAuth$'s core calculus by directly analyzing its new authenticated types, \citeauthor{atkey-ads}'s approach achieves security and correctness by using a \emph{parametricity} property~\citep{reynolds-parametricity} of OCaml's module system.
\citeauthor{atkey-ads} hints at how existing proofs of parametricity for type systems related to OCaml's module system could be used to prove security and correctness, but gives no formal proof.
Moreover, existing parametricity theorems cannot be directly applied, since establishing security and correctness properties require reasoning about hash collisions, which is not covered. 

In this paper, we provide for the first time a complete proof of security and correctness for generating authenticated data structures using a typed module system.
Moreover, we also address the second and third limitations of the $\lambdaAuth$ approach described above.
In particular, we prove the correctness of several optimizations for proof objects supported by $\lambdaAuth$'s compiler.
Additionally, we show how to prove that hand-written, optimized implementations of operations on an authenticated data structure can be safely linked with code that is generated automatically.
This allows a developer to combine the ease of automatic generation for most operations, while still being able to manually optimize certain operations that need to be as efficient as possible.

Our proof uses the technique of program-logic-based logical relations~\citep{logical-logical-relations} which has been used in recent years to prove properties about strong guarantees provided by a range of sophisticated type systems, including data race freedom provided by Rust~\citep{jung-rustbelt}, type soundness of an extension of Scala's core type system \cite{giarrusso-scala-dot}, expressive information-flow control types~\citep{iris-tini, seloc}, and program refinement \citep{iris-logrel-journal, reloc, clutch, tassarotti-refinement, timany-runst, timany-continuations}.
With this technique, to prove that a type system guarantees a particular property, one starts by constructing a program logic, typically a variant of Hoare logic~\citep{hoare-logic}, that is expressive enough for proving that programs have the property in question. 
Next, one defines a \emph{semantic model} of the type system, in which the meaning of a type is defined in terms of Hoare triples in the program logic.
Using the rules of the logic, one proves that this model is sound by showing that whenever a program $e$ is well-typed according to the rules of the type system, a corresponding Hoare triple holds for $e$.
Such triples then guarantee that all well-typed programs have the desired property.

In our application of this technique, we construct a program logic called \thelogic for relational reasoning about programs that make use of collision-resistant hash functions~(\cref{sec:logic}).
Then, we construct two semantic models of a type system that can encode the module-based construction proposed by \citeauthor{atkey-ads}.
The first model captures the security property of authenticated data structures~(\cref{sec:logrel-sec}), while the second captures the correctness property~(\cref{sec:logrel-correctness}).
Using these models, we then prove the correctness of several of the optimizations implemented in the $\lambdaAuth$ compiler~(\cref{sec:optimizations}).
Next, we show how an optimized, manual implementation of retrieval operations for a Merkle tree can be soundly linked with automatically generated code for other operations on this data structure~(\cref{sec:semantic-proofs}).
Finally, we compare our approach to prior work on verifying the correctness of authenticated data structures (\cref{sec:work}).

All of the results in this paper have machine-checked proofs of correctness carried out with the Coq proof assistant~\citep{coq} using the Iris program logic framework~\citep{iris-journal}.
The complete proof development is available in the supplementary material.

%%% Local Variables:
%%% mode: latex
%%% TeX-master: "ccs25-auth-full-version"
%%% End:

%% file: background.tex
\section{Background}
\label{sec:background}

This section begins by describing \citeauthor{atkey-ads}'s module-based construction of authenticated data types.
Next, we define the formal calculus and type system that we use to encode this construction.

\subsection{Authentikit}
\label{sec:background:authentikit}

\citeauthor{atkey-ads} makes the observation that it is possible to implement the functionality offered by the \lambdaAuth programming language as a library \emph{within} a general-purpose programming language with sufficiently powerful abstraction facilities, such as OCaml.
The key idea is to express authenticated computations using an abstract monad that, depending on its instantiation, will either construct or consume proofs alongside the computation.
The abstract interface of \citeauthor{atkey-ads}'s library, called Authentikit, is given by the OCaml module signature shown in \cref{fig:authentikit:ocaml}.

\begin{figure*}[ht!]
\centering
\begin{subfigure}[t]{0.47\linewidth}
\begin{lstlisting}[language=ocaml]
module type AUTHENTIKIT = sig
  type 'a auth
  type 'a m
  val return : 'a -> 'a m
  val bind   : 'a m -> ('a -> 'b m) -> 'b m

  module Authenticatable : sig
    type 'a evi
    val auth   : 'a auth evi
    val pair   : 'a evi -> 'b evi -> ('a * 'b) evi
    val sum    : 'a evi -> 'b evi ->
                 [`left of 'a | `right of 'b] evi
    val string : string evi
    val int    : int evi
  end

  val auth   : 'a Authenticatable.evi ->
               'a -> 'a auth
  val unauth : 'a Authenticatable.evi ->
               'a auth -> 'a m
end
\end{lstlisting}
\caption{Module signature in OCaml.}
\label{fig:authentikit:ocaml}
\end{subfigure}
\begin{subfigure}[t]{0.47\linewidth}
  \small
  \begin{align*}
    \tAuthentikit
    &\eqdef{} \Exists \tauth, \tauthcomp : \karr{\ktype}{\ktype}. \tAuthentikitfunc~\tauth~\tauthcomp \\
    \tAuthentikitfunc
    &\eqdef{} \lambda \tauth,  \tauthcomp : \karr{\ktype}{\ktype} \ldotp \\
    &\hspace{2em}
      \begin{aligned}
        &\left(\All \tvar : \ktype. \tvar \to \tauthcomp~\tvar\right)  \times{} \\
        &\left(\All \tvar, \tvarB : \ktype . \tauthcomp~\tvar \to (\tvar \to \tauthcomp~\tvarB) \to \tauthcomp~\tvarB\right) \times{} \\
        &\tAuthenticatable
      \end{aligned} \\
    \tAuthenticatable
    &\eqdef{} \Exists \tevi : \karr{\ktype}{\ktype} . {} \\
    &\hspace{2em}
      \begin{aligned}
        &\left(\All \tvar : \ktype . \tevi~(\tauth~\tvar)\right) \times \\
        &\left(\All \tvar, \tvarB : \ktype . \tevi~\tvar \to \tevi~\tvarB \to \tevi~(\tvar \times \tvarB) \right) \times \\
        &\left(\All \tvar, \tvarB : \ktype . \tevi~\tvar \to \tevi~\tvarB \to \tevi~(\tvar + \tvarB) \right) \times \\
        &\left(\tevi~\tstring\right) \times{} \\
        &\left(\tevi~\tint\right) \times{} \\
        &\ghostcode{\left(\All f : \karr{\ktype}{\ktype} .  \tevi~(f (\mu_{\ktype} f)) \to \tevi~(\mu_{\ktype} f)\right) \times{}} \\
        &\left(\All \tvar : \ktype . \tevi~\tvar \to \tvar \to \tauth~\tvar\right) \times{} \\
        &\left(\All \tvar : \ktype . \tevi~\tvar \to \tauth~\tvar \to \tauthcomp~\tvar\right)
      \end{aligned}
  \end{align*}
  \vspace{-0.5em}
  \caption{Module signature as a type in \thelang.}
  \label{fig:authentikit:thelang}
\end{subfigure}
\caption{Authentikit type signatures.}
\label{fig:authentikit}
\end{figure*}

First, the signature postulates the existence of a type constructor \texttt{auth} that represents the type of authenticated values.\footnote{This constructor corresponds to the authenticated type constructor ``$\authty$'' found in \lambdaAuth.}
Note that \texttt{auth} is abstract and the interface does not commit to a particular choice for how authenticated types are represented.

Second, the interface requires programmers to structure authenticated computations using an \emph{authenticated computation monad} given by the (abstract) type constructor \texttt{m} and the two operations \texttt{return} and \texttt{bind}.
As usual with monadic programming, the \texttt{bind} operation allows multiple steps of authenticated computation to be sequenced together.
As we will see, forcing programs to be structured with these primitives allows the library to perform operations for handling and checking proofs between steps of execution.

Third, as the implementation will use hash functions for authenticating data, the interface requires a way to show that the data we want to authenticate can be hashed.
In particular, there must be a way to serialize the value to a string.
For example, OCaml has first class functions as values, but there is no way to serialize these functions from within the language, so we cannot use hashes to authenticate a function.
Authentikit requires the programmer to build explicit serialization functions using the combinators in the \texttt{Authenticatable} submodule.
A term of type \texttt{t evi} is a serialization function for data of type \texttt{t}.

Finally, there are two functions \texttt{auth} and \texttt{unauth}.
The \texttt{auth} function produces authenticated values, and, given an authenticated value, the \texttt{unauth} function returns an authenticated computation that ``unwraps''  the authenticated value.
The two functions correspond directly to two built-in operations of the same names found in \lambdaAuth.
As we will see, the behavior of these two functions differs significantly between the verifier and prover implementations of the Authentikit interface.

\paragraph{Merkle Trees}
Before we discuss how the Authentikit interface can be implemented, consider an implementation of basic Merke trees using Authentikit shown in \cref{fig:merkle}.

\begin{figure}
  \centering
\begin{lstlisting}[language=ocaml]
module Merkle = functor (K : AUTHENTIKIT) -> struct
  open K

  type path = [`L | `R] list
  type tree = [`leaf of string |
               `node of tree auth * tree auth]

  let tree_evi : tree Authenticatable.evi =
    Authenticatable.(sum string (pair auth auth))

  let make_leaf (s : string) : tree auth =
    auth tree_evi (`leaf s)
  let make_branch (l r : tree auth) : tree auth =
    auth tree_evi (`node (l,r))

  let rec retrieve (p : path) (t : tree auth)
    : string option m =
    bind (unauth tree_evi t) (fun t ->
      match p, t with
      | [], `leaf s -> return (Some s)
      | `L::p, `node (l,_) -> retrieve p l
      | `R::p, `node (_,r) -> retrieve p r
      | _, _ -> return None)

  let rec update (* ... *) = (* ... *)
end
\end{lstlisting}
  \caption{Merkle trees implemented using Authentikit.}
  \label{fig:merkle}
\end{figure}

The \texttt{Merkle} module is parameterized by an implementation of the Authentikit interface.
It defines a type for paths and a type for (authenticated) trees as well as operations for constructing, querying, and updating trees.
Notice how the type signatures matches the interface one would expect for ordinary binary trees, with the addition of authentication annotations using the \texttt{auth} and \texttt{m} constructors.
For example, rather than returning a tree, \texttt{make\_leaf} returns an authenticated tree;
and rather than taking a path and a tree as input and returning the result of a query, \texttt{retrieve} takes a path and an authenticated tree as input and returns the result of the query in the authenticated computation monad.

The function \texttt{tree\_evi} defines a serializer for trees using the serialization primitives from the \texttt{Authenticatable} submodule.
The \texttt{make\_leaf} and \texttt{make\_branch} functions are implemented using \texttt{auth} and the corresponding data constructors.
Finally, the \texttt{retrieve} operation is just an ordinary binary tree traversal but with a few annotations: trees are first unwrapped using \texttt{unauth} and the computation is tracked using \texttt{bind} and \texttt{return}.
The \texttt{update} function is defined in a similar way.

% The \texttt{Merkle} module defines operations on Merkle trees as authenticated computations parameterized by the Authentikit interface.
Because the \texttt{Merkle} module is parameterized by \texttt{Authentikit}, to run a computation, we first have to instantiate the module with an implementation of \texttt{Authentikit}.
\Cref{fig:authentikit-impl} shows three such implementations that will give rise to the prover, verifier, and ``ideal'' semantics of the data structure.

\paragraph{Prover Implementation}

The \texttt{Prover} module shown in \cref{fig:authentikit:prover} implements the Authentikit module signature.
The prover views authenticated values as a pair of an underlying value and a hash of its string representation.
Authenticated computations are thunks that return a pair of a proof and a result.\footnote{We deviate slightly from \citeauthor{atkey-ads}'s original presentation by implementating authenticated computations as thunks. This is necessary for the correctness theorems to hold in the presence of side effects.}
Proofs are represented as lists of strings.
The \texttt{bind} operation appends together the proofs that are generated by each step of the computation.

\begin{figure*}
  \centering
  \begin{minipage}[t]{0.49\linewidth}
    \begin{subfigure}[t]{1\linewidth}
\begin{lstlisting}[language=ocaml]
type proof = string list

module Prover : AUTHENTIKIT =
  type 'a auth = 'a * string
  type 'a m = () -> proof * 'a
  let return a () = ([], a)
  let bind c f =
    let (prf,  a) = c () in
    let (prf', b) = f a () in
    (prf @ prf', b)

  module Authenticatable = struct
    type 'a evidence = 'a -> string
    let auth (_, h) = h
    (* ... *)
  end

  let auth evi a = (a, hash (evi a))
  let unauth evi (a, _) () = ([evi a], a)
  let run m = m ()
end
\end{lstlisting}
      \caption{Prover. \textnormal{$\authentikit_{P}$} denotes the corresponding term in \thelang. \linebreak}
      \label{fig:authentikit:prover}
    \end{subfigure}

  \begin{subfigure}[t]{1\linewidth}
\begin{lstlisting}[language=ocaml]
module Ideal : AUTHENTIKIT = struct
  type 'a auth = 'a
  type 'a m = () -> 'a

  let return a () = a
  let bind a f () = f (a ()) ()
  (* .. *)
  let auth _ a = a
  let unauth _ a () = a
end
 \end{lstlisting}
    \caption{Ideal. \textnormal{$\authentikit_{I}$} denotes the corresponding term in \thelang.}
    \label{fig:authentikit:ideal}
  \end{subfigure}
\end{minipage}
\hfill
\begin{minipage}[t]{0.49\linewidth}
  \begin{subfigure}[t]{1\linewidth}
\begin{lstlisting}[language=ocaml]
type proof = string list

module Verifier : AUTHENTIKIT =
  type 'a auth = string
  type 'a m =
    proof -> [`Ok of proof * 'a | `ProofFailure ]

  let return a prf = `Ok (prf, a)

  let bind c f prf =
    match c prf with
    | `ProofFailure -> `ProofFailure
    | `Ok (prf', a) -> f a prf'

  module Authenticatable = struct
    type 'a evidence =
      { serialize : 'a -> string;
        deserialize : string -> 'a option }
    (* ... *)
  end

  let auth evi a = hash (evi.serialize a)

  let unauth evi h prf =
    match prf with
    | p :: ps when hash p = h ->
      match evi.deserialize p with
      | None   -> 'ProofFailure
      | Some a -> `Ok (ps, a)
    | _ -> 'ProofFailure

  let run cf prf =
    match c prf with
    | `Ok (_, a) -> a
    | _ -> failwith "Proof failure"
end
\end{lstlisting}
    \caption{Verifier. \textnormal{$\authentikit_{V}$} denotes the corresponding term in \thelang.}
    \label{fig:authentikit:verifier}
  \end{subfigure}
\end{minipage}
\caption{Three realizations of the Authentikit interface in OCaml..}
\label{fig:authentikit-impl}
\end{figure*}

Authenticatable values are, from the perspective of the prover, values for which there exist a serialization function.
The combinators are implemented following a consistent serialization scheme; we omit the details but discuss the requirements in \cref{sec:logrel-sec}.
In the case of \texttt{auth}, the serialization of an authenticated value is its hash code, the ``digest'', not the underlying value.
This means that, for example, the string representation of an authenticated tree is the root hash.
Finally, the \texttt{auth} function pairs an input with its hashed string representation.
The \texttt{unauth} function returns the underlying value and produces a proof containing the serialization of the value.

When we instantiate the \texttt{Merkle} module with the \texttt{Prover} module, each recursive call in \texttt{retrieve} will generate a singleton proof from the \texttt{unauth} invocation used to access the current node.
These proofs are then appended together through the \texttt{bind} operations to produce an overall proof for the complete call to \texttt{retrieve}.
\Cref{fig:merkle-tree-prover} depicts a Merkle tree from the perspective of the prover.
Calling \texttt{retrieve} on the authenticated tree $(t_{0}, h_{0})$ and the path $[\textsf{R}, \textsf{L}]$ produces the proof $[(h_{1}, h_{2}), (h_{5}, h_{6}), s_{5}]$.
There is some redundancy in the proof that is generated during a call to \texttt{retrieve}, as compared to a typical, manual implementation of a Merkle tree; \citet{miller-ads} noted a similar redundancy in the code generated for \texttt{retrieve} in $\lambdaAuth$.
We will revisit this in \cref{sec:optimizations}, where we discuss general optimizations that may reduce the proof size, and in \cref{sec:semantic-proofs}, where we verify a manual implementation of \texttt{retrieve} that generates the optimal proof.

\newcommand{\treehash}{\text{hash}}
\newcommand{\treenode}{\text{node}}
\newcommand{\treeleaf}{\text{leaf}}
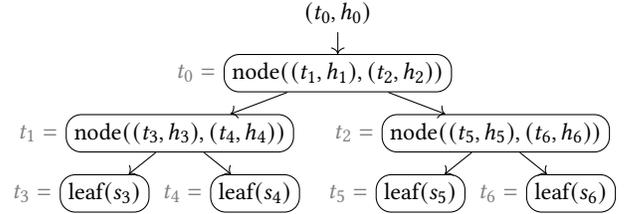
\begin{figure}
  \centering
  \begin{tikzpicture}
    [level distance=8mm,
    level/.style={sibling distance=42mm},
    edge from parent/.style={draw,->}]
    \node {$(t_0, h_0)$}
    child {
      node[draw, rounded corners=2mm, label=left:{\color{gray}$t_{0}=$}]
      {$\treenode((t_{1}, h_{1}), (t_{2}, h_{2}))$} edge from parent
      child[level/.style={sibling distance=20mm}]{
        node[draw, rounded corners=2mm, label=left:{\color{gray}$t_{1}=$}]
        {$\treenode((t_{3}, h_{3}), (t_{4}, h_{4}))$}
        child {node[draw, label=left:{\color{gray}$t_{3}=$}, rounded corners=2mm] {$\treeleaf(s_{3})$}}
        child {node[draw, label=left:{\color{gray}$t_{4}=$}, rounded corners=2mm] {$\treeleaf(s_{4})$}}
      }
      child[level/.style={sibling distance=20mm}]{
        node[draw, rounded corners=2mm, label=left:{\color{gray}$t_{2}=$}]
        {$\treenode((t_{5}, h_{5}), (t_{6}, h_{6}))$}
        child {node[draw, label=left:{\color{gray}$t_{5}=$}, rounded corners=2mm] {$\treeleaf(s_{5})$}}
        child {node[draw, label=left:{\color{gray}$t_{6}=$}, rounded corners=2mm] {$\treeleaf(s_{6})$}}
      };
    };
  \end{tikzpicture}
  \caption{The prover view of a Merkle tree where $h_{i}$ is the hash of $t_{i}$.
    The hash of a node is uniquely determined by the hashes of its children, \eg{}, $h_{2}$ is derived from $h_{5}$ and $h_{6}$.}
  \label{fig:merkle-tree-prover}
\end{figure}

\paragraph{Verifier Implementation}
The module \texttt{Verifier} in \cref{fig:authentikit:verifier} also implements the Authentikit signature.
The verifier views authenticated values as strings, \ie{}, just the hash code of the corresponding value.
Authenticated computations, meanwhile, are functions that take proofs as input and either (1) return a value and a left-over proof, or (2) indicate failure for an invalid proof.

Authenticatable values are, from the perspective of the verifier, values for which there exists both a serialization function and a deserialization function.
The combinators are implemented following the same serialization scheme as the prover so that serialization followed by deserialization for authenticated values is the identity.

To create authenticated values, the \texttt{auth} function serializes and hashes the value.
For instance, the Merkle tree shown in \cref{fig:merkle-tree-prover} is from the verifier's perspective just the root hash $h_{0}$.
The \texttt{unauth} function, however, is the primary workhorse where the proof checking happens.
It receives a hash code and proof as input and checks that the hash code of the first item in the proof list matches.
If so, the proof item is deserialized and the result is returned together with the remaining proof.
If the proof list is empty, if the hash code does not match, or if the deserialization fails, then the verifier returns \texttt{\textasciigrave ProofFailure}.

Instantiating the \texttt{Merkle} module with the \texttt{Verifier} module yields an implementation of \texttt{retrieve} that, in symmetry with the prover, consumes and checks a proof item against a hash code for every node it encounters as it descends.

\paragraph{Ideal Implementation}

The module \texttt{Ideal} shown in \cref{fig:authentikit:ideal} implements the Authentikit module signature with the type constructors and operations as the identity.
The module is not intended to be executed but serves as a specification device to say how the data structure functionally should behave.
In particular, instantiating \texttt{Merkle} with the \texttt{Ideal} module yields an implementation of ordinary (non-authenticated) binary trees and, \eg{}, the ideal perspective of the Merkle tree shown in \cref{fig:merkle-tree-prover} is the corresponding binary tree where all the hash codes are erased.

\paragraph{Security and Correctness}
Now that we have seen the implementation of the Authentikit library, let us consider at a high level why this library has the security and correctness properties described in \cref{sec:introduction}.
Consider the security property and look at what happens when \texttt{Merkle} is instantiated with the \texttt{Verifier} module and the \texttt{Ideal} module.
We can compare how the \texttt{Verifier} version of \texttt{retrieve} runs when its input tree argument is a hash corresponding to the root of a tree given as input to the \texttt{Ideal} version of \texttt{retrieve}.
If the \texttt{unauth} function in the \texttt{Verifier} code does not return \texttt{\textasciigrave ProofFailure}, then the value it returns must have the same hash as the value the \texttt{Ideal} version is using. Thus, either these values are equal, or we have exhibited a hash collision.
Since the implementation of \texttt{retrieve} is the same outside of the code from the Authentikit module, this means that, absent a hash collision or a proof failure, the overall result of the retrieve operations matches.
Similarly, for the correctness property, when the verifier is given as input the proof produced by the prover, each call to \texttt{unauth} in \texttt{retrieve} on the verifier side should return the same node value that the prover used in the corresponding step.

In the end, it is not so challenging to prove that one \emph{particular} client of the Authentikit library, like \texttt{Merkle}, has the intended security and correctness properties.
However, our goal is to prove that \emph{any} well-typed client of the library has these properties.
This is much more challenging to show because it requires somehow exploiting the fact that any well-typed client must use the underlying Authentikit operations in a ``generic way'', as enforced by the type system, so that we can relate what happens when the different implementations of these operations are plugged into a client.

\subsection[The F-omega-mu-ref language]{The \smash{\thelang{}} Language}
\label{sec:background:thelang}

\begin{figure*}
\begin{minipage}[t]{0.5\textwidth}
\begin{align*}
      &  \text{(values)}
    & \val
    \bnfdef{}
    & % \TT \ALT
       % b \in \bool \ALT
       % z \in \integer \ALT
       % s \in \String \ALT
       % \loc \in \Loc \ALT
       % (\val,\val) \ALT
       % \Inl \val  \ALT
       % \Inr \val \ALT
       \ldots \ALT
       \Rec f \var = \expr \ALT
       \Lambda \expr \ALT
       \Pack \val
    \\
    &  \text{(terms)}
    & \expr
    \bnfdef{}
    & \ldots \ALT
       \val \ALT
       \var % \in \Var
       \ALT
       % \HLOp_{1} \expr \ALT
       % \expr \HLOp_{2} \expr \ALT
       \expr~\expr \ALT
       % \If \expr then \expr \Else \expr \ALT
       % \Fst \expr \ALT
       % \Snd \expr \ALT
       % \Case~\expr~\expr~\expr \ALT
       \Alloc \expr \ALT
       \deref \expr \ALT
       \expr \gets \expr \ALT
    \fold \expr \ALT
       \unfold \expr \ALT
    \\ & & 
    &
       \expr~\tapp \ALT
    \Pack \expr \ALT
       \Unpack \expr as \var in \expr \ALT
       \Hash \expr
\end{align*}
\end{minipage}%
\begin{minipage}[t]{0.5\textwidth}
\begin{align*}
  &  \text{(kinds)}
  & \kind
  \bnfdef{}
  & \ktype \ALT
     \karr{\kind}{\kind}
  \\
  &  \text{(types)}
  & \type
  \bnfdef{}
  & \tvar \ALT
     \lambda \tvar : \kind \ldotp \type \ALT
     \type~\type \ALT
     \tconstr
  \\
  & \text{(constructors)}
  & \tconstr
  \bnfdef{}
  & \ldots \ALT
     \times \ALT
     + \ALT
     \to \ALT
     \tref  \ALT
     \forall_{\kind} \ALT
     \exists_{\kind} \ALT
     \mu_{\kind}
\end{align*}
\end{minipage}
\caption{Excerpt of the grammar of syntax and types for $\thelang$.}
\label{fig:excerpt-lang-grammar}
\end{figure*}

As a first step toward proving that authenticated data structures implemented using Authentikit satisfy the intended security and correctness properties, we need a formal model of how the module type system works in a language like OCaml.
Although there have been many proposals for models of Standard ML and OCaml-like module systems, in this work we use an approach based on translating modules into terms in a variant of \fomega~\citep{girard-f-omega}.
Specifically, we use \thelang, a higher-order programming language with higher-order references and a type system with polymorphism, abstract data types, iso-recursive types, and type abstraction in the style of \fomega.
An excerpt of the syntax of \thelang is given in \Cref{fig:excerpt-lang-grammar}, while \Cref{app:language} shows the full syntax and typing rules.

The term language is mostly standard, with the addition of a primitive $\Hash$ operation.
Note that there are no types in terms: we write $\Lambda \expr$ for type abstraction, $\expr~\tapp$ for type application, and $\Pack~\val$ and $\Unpack \expr_{1} as \var in \expr_{2}$ for formation and elimination of abstract data types.
The operations $\fold \expr$ and $\unfold \expr$ are the special term constructs for iso-recursive types.
$\Alloc~\expr$ allocates a new reference, $\deref \expr$ dereferences the location $\expr$ evaluates to, and $\expr_{1} \gets \expr_{2}$ assigns the result of evaluating $\expr_{2}$ to the location that $\expr_{1}$ evaluates to.
We write $\lambda \var \ldotp \expr$ to mean $\Rec {\_} \var = \expr$, $\Let \var = \expr_{1} in \expr_{2}$ to mean $(\lambda \var \ldotp \expr_{2})~\expr_{1}$, and $\expr_{1} ; \expr_{2}$ to mean $\Let \_ = \expr_{1} in \expr_{2}$.

Types have a standard kind structure: the kind $\ktype$ for the kind of proper types and $\karr{\kind_{1}}{\kind_{2}}$ for constructors that given a type of kind $\kind_{1}$ produce a type of kind $\kind_{2}$.
Types are formed from type variables $\tvar$, type abstractions $\lambda \tvar : \kind \ldotp \type$, type applications $\type_{1}~\type_{2}$, and constructors $\tconstr$.
Constructors include unit, Booleans, integers, strings, products, disjoint sums, arrows, and references as well as universal, existential, and recursive quantifiers.
We write binary constructors using infix notation, \eg{}, $\type \to \typeB$ to mean $\to~\type~\typeB$, and we write $\All \tvar : \kind .
\type$ to mean $\forall_{\kind}~(\lambda \tvar : \kind \ldotp \type)$, and similarly for existential and recursive types.
The typing judgment $\Theta \mid \Gamma \vdash \expr : \type$ is mostly standard and assigns a type of kind $\ktype$ to a term in contexts $\Theta$ and $\Gamma$.
The context $\Theta$ assigns kinds to type variables and $\Gamma$ assigns types of kind $\ktype$ to term-level variables.

\Cref{fig:authentikit:thelang} shows the Authentikit OCaml module signature from \cref{fig:authentikit:ocaml} translated into the type $\tAuthentikit$ in \thelang.
The type is derived by applying the translation described by \citet{f-ing-module}.
Note that when defining the $\tAuthentikit$ type, we introduce an intermediate definition $\tAuthentikitfunc$ which will be useful for stating our security and correctness theorems.

While OCaml uses iso-recursion for nominal types such as variants and records, the polymorphic variants used in Authentikit are equi-recursive.
To avoid modeling both iso-recursive and equi-recursive types (which adds considerable complexity, see \eg{}, \cite{f-omega-equi-recursive}), we use iso-recursive types throughout and instead add an explicit % coercion
operation (shown in \ghostcode{gray}) for creating evidence of recursive types.

Throughout the rest of the paper, when presenting extended code snippets, we will continue to use OCaml syntax for readability as in \cref{fig:authentikit:ocaml}, however, in each case, all of our formal proofs are stated in terms of translations of these programs into \thelang.

%%% Local Variables:
%%% mode: latex
%%% TeX-master: "ccs25-auth-full-version"
%%% End:

%% file: logic.tex
\section{Collision-Free Reasoning in Separation Logic}\label{sec:logic}

As outlined in \cref{sec:introduction}, our first step in constructing a model of \thelang's type system is a program logic, \emph{Collision-Free Separation Logic} (\thelogic), that is expressive enough to state and prove our desired security and correctness properties.
\thelogic is built as an extension to the Iris program logic~\citep{iris-journal}, which is a modern variant of separation logic~\citep{reynolds-separation-logic}.
The key extension we add is a rule that internalizes the collision-resistance property of the cryptographic hash functions we use, allowing us to only consider execution traces that are \emph{collision-free} when proving a specification.

\thelogic's main program specification construct is a \emph{weakest precondition} assertion of the form $\wpre{\expr}{\Phi}$. 
In most separation logics with weakest preconditions, $\wpre{\expr}{\Phi}$ holds in a state $\sigma$ if, when executing $e$ in $\sigma$, execution of $e$ is \emph{safe} (meaning that the program never gets stuck or triggers an assertion failure), and upon termination, the resulting state will satisfy the postcondition $\Phi$.
In \thelogic, we weaken the meaning of $\wpre{\expr}{\Phi}$ to only require safety and postcondition-satisfaction for executions in which $e$ does not compute a hash collision.

\begin{figure}[t!]
  \centering
  \begin{align*}
    &\expr_{1} \purestep \expr_{2} \sep \wpre{\expr_{2}}{\pred} \proves \wpre{\expr_{1}}{\pred} && \ruletag{wp-pure} \\
    &\TRUE \proves \wpre{\Alloc \val}{\loc \ldotp \progheap{\loc}{\val}} && \ruletag{wp-alloc} \\
    &{\progheap{\loc}{\val}} \proves \wpre{\deref\loc}{ \val \ldotp \progheap{\loc}{\val}} && \ruletag{wp-load} \\
    &\progheap{\loc}{\val} \proves \wpre{\loc \gets \valB}{\_ \ldotp \progheap{\loc}{\valB} } && \ruletag{wp-store}  \\
    &\pred(\val) \proves \wpre{\val}{\pred} && \ruletag{wp-val} \\
    &\wpre{\expr}{\val .\, \wpre{\fillctx\lctx[\val]}{\pred}} \proves \wpre{\fillctx\lctx[\expr]}{\pred} && \ruletag{wp-bind} \\
    &(\All \val . \Psi(\val) \wand \pred(\val)) \sep \wpre{\expr}{\Psi} \proves \wpre{\expr}{\pred} && \ruletag{wp-wand}
    % \prop \sep \wpre{\expr}{\pred} &\proves \wpre{\expr}{v . \, \prop \sep \pred(v)} && \ruletag{wp-frame} \\
  \end{align*}
  \caption{Standard weakest precondition rules.
  }
  \label{fig:wp-rules}
\end{figure}

To state this precisely, we augment $\thelang$'s semantics to track a history of all hashes computed during execution. Assume a family of hash functions $(\hash_{i})_{i \in \HashFamilyIndex}$.
We consider an operational semantics with program states consisting of a heap (modeled as a finite map from locations to values), a \emph{history} of the strings that have been hashed during execution, and the current hash family index.
\[
  \pstate \in \State \eqdef{} (\Loc \fpfn \Val) \times \Set(\String) \times \HashFamilyIndex
\]
The operational semantics $(\cdot \to \cdot) \in (\Expr \times \State) \times (\Expr \times \State)$ extends the history when a program performs a $\Hash$ operation but is otherwise standard.
\[
  \langle\Hash s, (m, h, i)\rangle \to \langle\hash_{i}(s), (m, h \cup \{s\}, i)\rangle
\]
We define a \emph{collision-free step}
\begin{align*}
  \langle\expr, \pstate\rangle \cfstep \langle\expr', \pstate'\rangle \eqdef{}
  \langle\expr, \pstate\rangle \step \langle\expr', \pstate'\rangle \land \collisionFree(\pstate')
\end{align*}
where
\begin{align*}
  \collisionFree(m, h, i) &\eqdef{} \not\exists s_{1}, s_{2} \in h \ldotp \collision(s_{1}, s_{2}, i) \\
  \collision(s_{1}, s_{2}, i) &\eqdef{} s_{1} \neq s_{2} \wedge \hash_{i}(s_{1}) = \hash_{i}(s_{2}).
\end{align*}
We use $\cfsteps$ to denote the reflexive transitive closure of $\cfstep$ and take the predicate $\cfsafe(\expr)$ to mean that all collision-free executions of $\expr$ are safe.
The soundness theorem of \thelogic shown below only considers collision-free execution traces.
\begin{theorem}[Soundness]\label{thm:unary-adequacy}
  Let $\varphi$ be a first-order predicate. % over values
  If
  \begin{align*}
    \TRUE \proves \wpre{\expr}{\varphi}
  \end{align*}
  is derivable then $\cfsafe(\expr)$ and for all $i \in \HashFamilyIndex$, if $\langle\expr,(\emptyset, \emptyset, i)\rangle \cfsteps \langle\val, \pstate\rangle$ then $\varphi(\val)$ holds at the meta level.
\end{theorem}

\thelogic satisfies all the standard separation logic rules (see \cref{fig:wp-rules} for an excerpt) from the Iris program logic.
The rules use the separating conjunction connective, $P \sep Q$, which holds in some program state $\sigma$ if it is possible to decompose $\sigma$ into two disjoint pieces, $\sigma_1$ and $\sigma_2$, which satisfy $P$ and $Q$ respectively.
The separating implication $P \wand Q$ is a form of implication that is the right adjoint of $\sep$, in the sense that $P \sep (P \wand Q) \vdash Q$.
The points-to assertion $\progheap{\loc}{\val}$ holds in a state $\sigma$ if location $\loc$ in $\sigma$ stores the value $\val$.
In contrast to standard conjunction, $P \not\vdash P * P$ in general, since it may not be possible to split a state into two sub-pieces that each satisfy $P$.
Because separation logic assertions delineate a part of program state, assertions are often called \emph{resources}.

To reason in a collision-free compositional manner, \thelogic introduces two new resource assertions $\hashindex(i)$ and $\hashed(i, s)$.
The resource $\hashindex(i)$ records that $i$ is the hash family index and $\hashed(i, s)$ captures that $s$ can be found in the hash history and thus has been hashed using $\hash_{i}$ at some point during execution.

The rule \ruleref{wp-hash} shown below reflects the operational behavior of the $\Hash$ operation.
In the postcondition the $\hashed(i, s)$ resource is obtained, witnessing the fact that the string $s$ has in fact been hashed using $\hash_{i}$ during execution.
\[
  \inferrule[{\TirNameStyle{wp-hash}}]
  {}
  {\hashindex(i) \proves \wpre{\Hash s}{ \val \ldotp \val = \hash_{i}(s) \sep \hashed(i, s)} }
  {\label{wp-hash}}
\]

The $\hashed(i, s)$ resource is duplicable, \ie,
\[
  \hashed(i, s) \provesIff \hashed(i, s) \sep \hashed(i, s)
\]
and satisfies the rule \ruleref{hash-validity} which embodies reasoning without collisions: if two hashed strings witness a hash collision then the goal can trivially be discharged.
\begin{mathpar}
  \inferH{hash-validity}
  { \collision(s_1, s_2, i) }
  { \hashed(i, s_1) \sep \hashed(i, s_2) \proves \FALSE }
\end{mathpar}

\paragraph{Relational Collision-Free Reasoning}
So far, what we have seen is a \emph{unary} logic that allows us to prove properties of a single program.
However, our goal is to relate the behaviors of a prover, verifier, and ideal version of a program.
To do so, we need a \emph{relational} logic that will allow us to prove a correspondence between the behaviours of multiple programs.
We follow CaReSL~\citep{caresl} and construct a relational variant of \thelogic by encoding a second program as a resource $\spec(\expr)$.
The resource tracks a ``specification'' program which can be updated and progressed according to the operational semantics.
For example, the resource $\spec((\lambda \var \ldotp \expr_{1})~\val_{2})$ can be updated to $\spec(\subst{\expr_{1}}{\var}{\val_{2}})$, reflecting execution of a beta reduction.
Formally, this is specified as a \emph{view shift} implication~\citep{iris-journal}
$\spec((\lambda \var \ldotp \expr_{1})~\val_{2}) \vs \spec(\subst{\expr_{1}}{\val_{2}}{\var})$
in Iris. A view shift $\prop \vs \propB$ intuitively says that, given resources satisfying $\prop$, we can update our resources and the ``logical state'' to obtain resources satisfying $\propB$.
In particular,
\begin{mathpar}
  \infer
  {\propB \vdash \wpre{\expr}{\Phi}}
  {(\prop \vs \propB) \sep \prop \vdash \wpre{\expr}{\Phi}}
\end{mathpar}

The relational logic also comes with a points-to connective \mbox{$\specheap{\loc}{\val}$} that denotes ownership of the location $\loc$ and its contents $\val$ on the heap of the specification program.
For example, storing a value to a location in the specification program requires ownership of the points-to connective, as captured by the following rule:
\[
  \spec(\loc \gets \valB) \sep \specheap{\loc}{\val} \vs \spec(\TT) \sep \specheap{\loc}{\valB}
\]
in which, on the right hand side, we have $\spec(\TT)$ (reflecting that the store returned the unit value $\TT$), and the points-to assertion is updated to reflect the updated value of the location.
We refer to~\citet{reloc} and our Coq formalization for a detailed discussion on how the specification resources are defined.

A relational variant of the soundness theorem of \thelogic follows as a consequence from the resource construction and \cref{thm:unary-adequacy}.
\begin{corollary}[Relational Soundness]\label{thm:binary-adequacy}
  Let $\varphi$ be a first-order relation. %over values.
  If 
  \begin{align*}
    % \specCtx \sep
    \spec(\expr_{2}) \proves \wpre{\expr_{1}}{\val_{1} \ldotp \Exists \val_{2} . \spec(\val_{2}) \sep \varphi(\val_{1}, \val_{2})}
  \end{align*}
  is derivable then $\cfsafe(\expr)$ and for all $i \in \HashFamilyIndex$, if $\langle\expr_{1},(\emptyset, \emptyset, i)\rangle \cfsteps \langle\val_{1}, \pstate_{1}\rangle$
  there exists a value $\val_2$ and state $\pstate_{2}$ such that $\langle\expr_{2}, (\emptyset, \emptyset, i)\rangle \to^{\ast} \langle\val_{2}, \pstate_{2}\rangle$ and $\varphi(\val_{1}, \val_{2})$ holds at the meta level.
\end{corollary}

%%% Local Variables:
%%% mode: latex
%%% TeX-master: "ccs25-auth-full-version"
%%% End:

%% file: logrel.tex
\section{Security}\label{sec:logrel-sec}

In this section, we show that ADSs implemented using Authentikit have the security property described in \cref{sec:introduction}.
Formally stated, the security theorem looks as follows. We say a type $\tau$ is a \emph{primitive type} if it is either $\tunit$, $\tbool$, $\tint$, $\tstring$, $\type_{1} \times \type_{2}$, or $\type_{1} + \type_{2}$ where $\type_{1}$ and $\type_{2}$ are primitive types. 
\begin{theorem}[Security]\label{thm:security:syntactic}
  Let $\type$ be a primitive type.
  If
  \[
    \emptyset \mid \emptyset \proves \expr : \All \tauth, \tauthcomp : \karr{\ktype}{\ktype} . \tAuthentikitfunc~\tauth~\tauthcomp \to \tauthcomp~\type
  \]
  then for all proofs $p$ (a list of strings) and $i \in \HashFamilyIndex$,
  if
  \[\langle \run_{V}~(\expr~\tapp~\tapp~\authentikit_{V})~p, (\emptyset, \emptyset, i)\rangle \cfsteps \langle \Some\val, \pstate_1 \rangle\]
  then there exists a state $\pstate_{2}$ such that
  \[\langle \run_{I}~(\expr~\tapp~\tapp~\authentikit_{I}), (\emptyset, \emptyset, i)\rangle \steps \langle\val, \pstate_2\rangle\]
\end{theorem}
In prose, the theorem requires that $\expr$ is a well-typed function that takes an Authentikit implementation as an argument and returns an authenticated computation.
Then it says that if we instantiate $\expr$ with the verifier implementation $\authentikit_{V}$ and it accepts the proof $p$, it will return the same value as the ideal semantics given by instantiating and running $\expr$ with $\authentikit_{I}$.

The challenge in proving this theorem is that all we know about $\expr$ is that it has the type stated in the premise.
Intuitively, since $e$ has this type, it must use the operations provided by the Authentikit interface in a generic way and cannot violate the abstractions of the interface.
To prove the theorem formally, we must reason about these abstraction guarantees enforced by the type system.
A standard approach for reasoning about type abstraction is to construct a model using logical relations.
In this section, we construct such a model using \thelogic and use it to prove the theorem.

\subsection{Logical Relation for Security}\label{sec:sec:ref}

A logical-relations model provides an interpretation for a type system by describing the behaviors that all programs with a given type $\tau$ should have.
Constructing such a model ``directly'' for a highly expressive type system like that of $\thelang$ is challenging, but in recent years, the so-called ``logical'' approach to logical relations \cite{logical-logical-relations} has made this easier by defining the logical relation in terms of a highly expressive program logic.
This approach has been used with the Iris program logic for a wide range of languages with System F-like type systems, \eg{}, to prove program refinement \cite{reloc}.

To define a model of \thelang{}, we take a similar approach using the relational collision-free logic from \cref{sec:logic}, and adapt techniques from recent work of \citet{sieczkowski-gadt} to incorporate the higher kinds found in \thelang.
For completeness, the full model is defined in \cref{fig:bin-model}, though many of the technical details of the construction are not needed to understand the rest of our results. 

\begin{figure*}
  \centering
  \begin{minipage}[t]{0.48\textwidth}
    \begin{align*}
      \intertext{\textbf{Kind interpretation}}
      \Sem{\ktype}
      &\eqdef{} \Val \times \Val \to \piProp\\
      \Sem{\karr{\kind_{1}}{\kind_{2}}}
      &\eqdef{} \Sem{\kind_{1}} \tone \Sem{\kind_2}
        \intertext{\textbf{Type interpretation}}
        \Sem{\Theta \vdash \type : \kind}_{(\cdot)}
      &\; : \, \Sem{\Theta} \tone \Sem{\kind} \\
      \Sem{\Theta \vdash \tvar : \kind}_{\Delta}
      &\eqdef{} \Delta(\tvar) \\
      \Sem{\Theta \vdash \lambda \tvar \ldotp \type : \karr{\kind_{1}}{\kind_{2}}}_{\Delta}
      &\eqdef{} \lambda \relvar : \Sem{\kind_{1}} \ldotp \Sem{\Theta, \tvar : \kind_{1} \vdash \type : \kind_{2}}_{\Delta, \relvar} \\
      \Sem{\Theta \vdash \typeB~\type : \kind_{2}}_{\Delta}
      &\eqdef{} \Sem{\Theta \vdash \typeB : \karr{\kind_{1}}{\kind_{2}}}_{\Delta} \left( \Sem{\Theta \vdash \type : \kind_{1}}_{\Delta} \right) \\
      \Sem{\Theta \vdash c : \kind}_{\Delta}
      &\eqdef{} \Sem{c : \kind}
    \end{align*}
  \end{minipage}
  \begin{minipage}[t]{0.48\textwidth}
    \begin{align*}
      \intertext{\textbf{Term interpretation}}
      \SemE{\relvar}(\expr_{1}, \expr_{2})
      &\eqdef{} \All i, \lctx .
        \begin{aligned}[t]
          &\hashindex(i) \sep \spec(\fillctx \lctx [\expr_{2}]) \wand{} \\
          &\wpre{\expr_{1}}{\val_{1} \ldotp \Exists \val_{2} . \spec(\fillctx\lctx[\val_{2}]) \sep \relvar(\val_{1}, \val_{2}) }
        \end{aligned}
        \intertext{\textbf{Context interpretations}}
        \Sem{\Theta}
      &\eqdef{} \Pi_{\tvar : \kind \in \Theta} \Sem{\kind} \\
      \Sem{\Theta \vdash \Gamma}_{\Delta}(\vec{\val}, \vec{\valB})
      &\eqdef{}
        \All (\var_{i}, \type_{i}) \in \Gamma . \Sem{\Theta \vdash \type_{i} : \star}_{\Delta}(\val_{i}, \valB_{i})
        \intertext{\textbf{Logical relation}}
        \Theta \mid \Gamma \vDash \expr_{1} \preceq \expr_{2} : \type
          &\eqdef{} \All \Delta \in \Sem{\Theta}, \vec{\val_{1}}, \vec{\val_{2}} .
        \Sem{\Theta \vdash \Gamma}_{\Delta}(\vec{\val_{1}}, \vec{\val_{2}}) \wand \\
        &\qquad \SemE{\Sem{\Theta \vdash \type : \ktype}_{\Delta}}(\subst{\expr_{1}}{\Gamma}{\vec{\val_{1}}}, \subst{\expr_{2}}{\Gamma}{\vec{\val_{2}}})
    \end{align*}
  \end{minipage}
  \textbf{Constructor interpretation} \\
  \vspace{-1em}
  \raggedright
  \begin{minipage}[t][3.5em]{0.455\textwidth}
    \begin{align*}
      \Sem{\tunit : \ktype}
      &\eqdef{} \lambda (\val_{1}, \val_{2}) \ldotp \val_{1} = \val_{2} = \TT \\
      \Sem{\tbool : \ktype}
      &\eqdef{} \lambda (\val_{1}, \val_{2}) \ldotp \Exists b \in \bool . \val_{1} = \val_{2} = b \\
    \end{align*}
  \end{minipage}
  \begin{minipage}[t][3.5em]{0.3\textwidth}
    \begin{align*}
      \Sem{\tint : \ktype}
      &\eqdef{} \lambda (\val_{1}, \val_{2}) \ldotp \Exists z \in \integer . \val_{1} = \val_{2} = z \\
      \Sem{\tstring : \ktype}
      &\eqdef{} \lambda (\val_{1}, \val_{2}) \ldotp \Exists s \in \String . \val_{1} = \val_{2} = s
    \end{align*}
  \end{minipage}
  \begin{align*}
    \Sem{\times : \karr{\ktype}{\karr{\ktype}{\ktype}}}
    &\eqdef{} \lambda \relvar, \relvarB : \Sem{\ktype} \ldotp \lambda (\val_{1}, \val_{2}) \ldotp
      \Exists \valB_{1}, \valB_{2}, \valC_{1}, \valC_{2} .      \val_{1} = (\valB_{1}, \valC_{1}) \sep \val_{2} = (\valB_{2}, \valC_{2}) \sep
      \relvar(\valB_{1},\valB_{2}) \sep \relvarB(\valC_{1},\valC_{2}) \\
    \Sem{+ : \karr{\ktype}{\karr{\ktype}{\ktype}}}
    &\eqdef{} \lambda \relvar, \relvarB : \Sem{\ktype} \ldotp \lambda (\val_{1}, \val_{2}) \ldotp
      \Exists \valB_{1}, \valB_{2} .
      \left(\val_{1} = \Inl \valB_{1} \sep \val_{2} = \Inl \valB_{2} \sep \relvar(\valB_{1}, \valB_{2}\right)
      \lor \left(\val_{1} = \Inr \valB_{1} \sep \val_{2} = \Inr \valB_{2} \sep \relvarB(\valB_{1}, \valB_{2})\right) \\
    \Sem{\to : \karr{\ktype}{\karr{\ktype}{\ktype}}}
    &\eqdef{} \lambda \relvar, \relvarB : \Sem{\ktype} \ldotp \lambda (\val_{1}, \val_{2}) \ldotp
      \always \All \valB_{1}, \valB_{2} .
      \relvar(\valB_{1}, \valB_{2}) \wand \SemE{\relvarB}(\val_{1}~\valB_{1}, \val_{2}~\valB_{2}) \\
    \Sem{\tref : \karr{\ktype}{\ktype}}
    &\eqdef{} \lambda \relvar : \Sem{\ktype} \ldotp  \lambda (\val_{1}, \val_{2}) \ldotp
      \Exists \loc_{1}, \loc_{2} \in \Loc . \val_{1} = \loc_{1} \sep \val_{2} = \loc_{2} \sep
      \knowInv{}{\Exists \valB_{1}, \valB_{2} . \progheap{\loc_{1}}{\valB_{1}} \sep \specheap{\loc_{2}}{\valB_{2}} \sep \relvar(\valB_{1}, \valB_{2})} \\
    \Sem{\forall_{\kind} : \karr{(\karr{\kind}{\ktype})}{\ktype}}
    &\eqdef{} \lambda \relvar : \Sem{\karr{\kind}{\ktype}} \ldotp \lambda (\val_{1}, \val_{2}) \ldotp \All \relvarB \in \Sem{\kind} . \SemE{\relvar(\relvarB)}(\val_{1}~\tapp, \val_{2}~\tapp) \\
    \Sem{\exists_{\kind} : \karr{(\karr{\kind}{\ktype})}{\ktype}}
    &\eqdef{} \lambda \relvar : \Sem{\karr{\kind}{\ktype}} \ldotp \lambda (\val_{1}, \val_{2}) \ldotp \Exists \relvarB \in \Sem{\kind} . \relvar(\relvarB)(\val_{1}, \val_{2}) \\
    \Sem{\mu_{\kind} : \karr{(\karr{\kind}{\kind})}{\kind}}
    &\eqdef{} \lambda \relvar : \Sem{\karr{\kind}{\kind}} \ldotp \gfixp(F_{\kind}(\relvar))
      \hspace{3em} \text{where}~ F_{\kind}(\relvar)(f)  \eqdef{}
      \smash{
      \begin{cases}
        \lambda (\val_{1}, \val_{2}) \ldotp \later \relvar(f)(\val_{1}, \val_{2}) & \text{if}~\kind = \ktype \\
        \lambda \relvarB \in \Sem{\kind_{1}} \ldotp F_{\kind_2}(id)(\relvar(f)(\relvarB)) & \text{if}~\kind = \karr{\kind_{1}}{\kind_{2}}
      \end{cases}
      }
  \end{align*}
  \caption{Binary logical-relations model of \thelang.}
  \label{fig:bin-model}
\end{figure*}

First, the model defines an interpretation of types, $\Sem{\Theta \vdash \type : \kind}_{\Delta}$where $\Delta \in \Sem{\Theta}$ interprets the free type variables in $\type$.
Due to the higher-kinded nature of \thelang, the co-domain of this interpretation depends on the kind $\kind$ of $\type$.
The kind $\ktype$ of proper types is interpreted as binary relations in the logic.
Intuitively, $\Sem{\Theta \vdash \type : \ktype}_{\Delta}$ characterizes the set of pairs of closed values $(\val_{1}, \val_{2})$ of type $\type$ such that $\val_{1}$ \emph{refines} $\val_{2}$.\footnote{\thelogic is substructural, while the \thelang type system is not.
  To account for this, we consider relations defined as functions into $\piProp$, the type of \emph{persistent} propositions in \thelogic.
  We say $\prop$ is persistent if $\prop \proves \always \prop$ where $\always$ is the Iris \emph{persistence modality}.
}
The kind $\karr{\kind_{1}}{\kind_{2}}$ of constructors is interpreted as maps\footnote{More specifically, because the ambient logic of Iris is step-indexed, the maps also have to be \emph{non-expansive}, meaning that they map $n$-equivalent arguments to $n$-equivalent results. However, this is not important for an intuitive understanding of the model.} from $\Sem{\kind_{1}}$ to $\Sem{\kind_2}$.

The interpretation of types is defined by structural recursion on the type: type variables are interpreted by lookup in $\Delta$, type abstractions as maps, and type application as application of maps.
Constructors are interpreted using corresponding connectives in the logic in a standard way: \eg{}, functions are interpreted using (separating) implication taking related inputs to related outputs, universal types are interpreted using universal quantification, reference types are interpreted using the points-to connectives, and recursive types are interpreted using a guarded fixed point~\citep{nakano-guarded}.

Next, we define a term interpretation, $\SemE{\relvar}(\expr_{1}, \expr_{2})$, as a relational assertion in \thelogic.
This interpretation assertion is a separating implication that takes the $\hashindex(i)$ resource and the specification program executing $\expr_{2}$ as assumptions, and has as a conclusion a weakest precondition asserting that if $\expr_{1}$ reduces to some value $\val_{1}$, then there exists a corresponding execution of $\expr_{2}$ to a value $\val_{2}$ such that $\relvar(\val_{1}, \val_{2})$ holds.
Finally, the logical relation $\Theta \mid \Gamma \vDash \expr_{1} \preceq \expr_{2} : \type$ extends the interpretation to typed open terms by requiring that, after substituting in related values for each free variable, the resulting closed terms should be related according to $\SemENoArg$.

Using the rules of \thelogic, we next prove that the typing rules of \thelang are \emph{compatible} with the logical relation: for every typing rule, if we have a pair of related terms for every premise, then we also have  pair of related terms for the conclusion, \eg{}, in the case of function application, we have
\begin{mathpar}
  \infer
  {\Theta \mid \Gamma \vDash \expr_2 \preceq \expr_2' : \type_1 \to \type_2 \\
    \Theta \mid \Gamma \vDash \expr_1 \preceq \expr_1' : \type_1 }
  {\Theta \mid \Gamma \vDash \expr_2~\expr_1 \preceq \expr_2'~\expr_1' : \type_2 }
\end{mathpar}

Because such a compatibility lemma holds for every typing rule, the fundamental theorem of logical relations follows by induction on the typing derivation:
\begin{theorem}\label{thm:binary-fundamental}
  If $\Theta \mid \Gamma \vdash \expr : \type $
  then $\Theta \mid \Gamma \vDash \expr \preceq \expr : \type $.
\end{theorem}
That is, if $e$ is a well-typed term in $\thelang$, then $e$ is related to itself according to the logical relation.
In particular, if $e$ is a closed term, then the relational weakest precondition assertion given by $\SemENoArg$ holds for $e$.
Thus, just from knowing the type of $e$ we can deduce a ``free theorem'' automatically about $e$.

Taking a step back, there is nothing specific to authenticated data structures in this model.
The main changes, as compared to program-logic-based logical-relations models in prior work, is that (1) it includes support for higher kinds, and (2) the term interpretation $\SemENoArg$ uses \thelogic as the underlying program logic.
The fact that there is nothing specific to authenticated data types in the model is to be expected, since the idea behind the Authentikit library is that, unlike in $\lambdaAuth$, there is no need to extend OCaml with new special types for representing authenticated data.
Instead, security is ensured by the way that the library uses standard type system abstraction guarantees.
Thus, the ``security-specific'' work in showing \cref{thm:security:syntactic} lies in proving something about the implementations of Authentikit, which we turn to now.

\subsection{Security of Authentikit}
Let $\expr$ be a client of Authentikit satisfying the premises of \cref{thm:security:syntactic}.
Our goal in this section is to show the conclusion of this theorem by proving that when we instantiate $\expr$ with $\authentikit_{V}$ and $\authentikit_{I}$, the two different instantiations of $\expr$ are logically related according to the model of the previous section.
This will then imply that a relational weakest precondition in \thelogic holds between these two programs, so that our result will follow by applying \cref{thm:binary-adequacy}, the relational soundness theorem of \thelogic.

We start by applying the fundamental theorem of the logical relation to $\expr$ to get that it is related to itself at the type $\All \tauth, \tauthcomp : \karr{\ktype}{\ktype} . \tAuthentikitfunc~\tauth~\tauthcomp \to \tauthcomp~\type$.
Unfolding the definition of the logical relation as given in \cref{fig:bin-model}, it says that for any choice of relational interpretation of the quantified constructors $\tauth$ and $\tauthcomp$, if we apply $\expr$
to arguments that are related according to $\Sem{\tAuthentikitfunc~\tauth~\tauthcomp}_{\Delta}$, the results will be related according to $\SemE{\Sem{\tauthcomp~\tau}_{\Delta}}$ (where $\Delta$ maps $\tauth$ and $\tauthcomp$ to the selected interpretations).
In particular, we define interpretations of $\tauth$ and $\tauthcomp$ so that we can show the verifier and ideal implementation of Authentikit are related, \ie{}, we show
\[
  \Sem{\tAuthentikitfunc~\tauth~\tauthcomp}_{\Delta}(\authentikit_{V}, \authentikit_{I})
\]

For the interpretation $\Sem{\tauth} : \Sem{\ktype} \to \Sem{\ktype}$ of the constructor for authenticated values, we choose the following:
\begin{align*}
  \Sem{\tauth}(\relvar)(\val_{1}, \val_{2})
  &\eqdef{}  \Exists a, t, i .
    \begin{aligned}[t]
      & \val_{1} = \hash_{i}(\ser_{t}(a)) \sep \relvar(a, \val_{2}) \sep{} \\
      & \hashindex(i) \sep \hashed(i, \ser_{t}(a))
    \end{aligned}
\end{align*}
A pair of values $(\val_{1}, \val_{2})$ inhabits this relation if $\val_{1}$ (the verifier side) is the hash of the serialization of some value $a$ such that $\relvar(a, \val_{2})$.
where $\relvar$ is the relation that characterizes the type of $a$.
The two additional resources record that the hash function family used is $\hash_{i}$ and that the serialization of $a$ was hashed during execution.

The (meta-level) partial function $\ser_{t} : \Val \pfn \String$ maps values of type $t \in \{\tstring, \tint, t \times t, t + t\} $ to strings according to some serialization scheme for $t$.
Two crucial properties of the serialization strategy applied throughout the Authentikit library is injectivity and uniqueness: if $\ser_{t_{1}}(\val_{1}) = \ser_{t_{2}}(\val_{2})$ then $\val_{1} = \val_{2}$; and $\ser_{t_{1}}(\val) = \ser_{t_{2}}(\val)$ for all $t_{1}$ and $t_{2}$.

The interpretation $\Sem{\tauthcomp} : \Sem{\ktype} \to \Sem{\ktype}$ of the constructor for authenticated computations looks as follows.
\begin{align*}
  &\Sem{\tauthcomp}(\relvar)(\val_{1}, \val_{2})
    \eqdef{}
    \All \lctx, \valB, p .  \spec (\fillctx \lctx [\val_{2}~\TT]) \sep \isproof(w, p) \wand \\
  & \quad
    \wpre
    {\val_{1}~\valB}
    {
    \begin{aligned}
    &u_{1} \ldotp u_{1} = \None \lor{} \\
    &\quad \left(
    \begin{aligned}
      &\Exists a_{1}, a_{2}, w', p' . u_{1} = \Some~(w', a_{1}) \sep{} \\
      &\quad \spec (\fillctx \lctx [a_{2}]) \sep \isproof(w', p') \sep \relvar(a_{1}, a_{2})
    \end{aligned}
    \right)
    \end{aligned}
    }
\end{align*}
The specification says that the verifier $\val_{1}$, when applied to a value $w$ corresponding to the proof stream $p$, returns an option value indicating whether the proof was accepted or not.
If the proof is accepted, the verifier and the ideal return $(w', a_{1})$ and $a_{2}$, respectively, where $\relvar(a_{1}, a_{2})$ holds and $w'$ is a value corresponding to the remaining unconsumed proof stream $p'$.

We continue the proof by setting $\Delta = [\tauth \mapsto \Sem{\tauth}, \tauthcomp \mapsto \Sem{\tauthcomp}]$ and show that $\authentikit_{V}$ and $\authentikit_{I}$ are related component by component.
For the \authenticatable sub-module we pick an interpretation of the $\tevi$ constructor where $\Sem{\tevi}(\relvar)(\val_{1}, \val_{2})$ requires $\val_{1}$ to be a pair of a serializer and a deserializer that behaves as expected when applied to values inhabiting $\relvar$ according to the serialization scheme.

The most interesting case is $\unauth$, where proof checking happens and all parts of the model come together, \ie{}, we show
\[
  \Sem{\All \tvar : \ktype . \tevi~\tvar \to \tauth~\tvar \to \tauthcomp~\tvar}_{\Delta'}(\unauth_{V}, \unauth_{I})
\]
where $\Delta' = \Delta[\tevi \mapsto \Sem{\tevi}]$.
By unfolding the definition, we are to show that given $\Sem{\tevi}(\relvar)(v_{1}, v_{2})$ and $\Sem{\tauth}(\relvar)(w_{1}, w_{2})$ then $\SemE{\Sem{\tauthcomp}(\relvar)}(\unauth_{V}~v_{1}~w_{1},\unauth_{I}~v_{2}~w_{2})$.
By the interpretation of $\tauthcomp$, there is some proof $p$ which corresponds to $w_1$.
If $p$ is empty, the verifier returns $\None$ and we are done.
If $p = s :: p'$ then we continue using \ruleref{wp-hash}, obtaining the resource $\hashed(s)$.
Since $\Sem{\tauth}(\relvar)(w_{1}, w_{2})$ we know $w_{1} = \hash_{i}(\ser_{t}(a))$ and we have $\hashed(\ser_{t}(a))$ and $\relvar(a, w_{2})$ for some $t$ and $a$.
The next step of the verifier is thus the conditional test $\hash_{i}(s) = \hash_{i}(\ser_{t}(a))$.
If the test fails, the verifier returns $\None$ and we are done.
If the test succeeds then either (1) we have encountered a collision, in which case we conclude using \ruleref{hash-validity}, or (2) $s$ is equal to $\ser_{t}(a)$ which means $s$ is a valid serialization of $a$ and deserialization succeeds, so we are done.

At this point, we have shown
\[\SemE{\Sem{\tauthcomp~\tau}_{\Delta}}(\expr~\tapp~\tapp~\authentikit_{V}, \expr~\tapp~\tapp~\authentikit_{I})\]
giving us a relational weakest precondition for the two instantiations of the client $\expr$.
The last step of our security theorem is to prove, using this weakest precondition, that when we execute the verifier and prover with $\run_{V}$ and $\run_{I}$, the results match the conclusion of \cref{thm:security:syntactic}.
We prove this as a separate lemma about $\run_{V}$ and $\run_{I}$ by establishing a weakest precondition in \thelogic, and then applying \cref{thm:binary-adequacy}, to get
\begin{lemma}[Security, semantic]\label{thm:security:sem}
  Let $\varphi$ be a first-order relation. If
  $\SemE{\Sem{\textup{\tauthcomp}}~\varphi}(\expr_{1}, \expr_{2})$ % then for all proofs $p$ and $i \in \HashFamilyIndex$
  then for all proofs $p$, if
  $\langle \run_{V}~\expr_{1}~p, (\emptyset, \emptyset, i)\rangle \cfsteps \langle \Some\val_{1}, \pstate_1 \rangle$
  then $\langle \run_{I}~\expr_{2}, (\emptyset, \emptyset, i)\rangle \steps \langle\val_{2}, \pstate_2\rangle$
  such that $\varphi(\val_{1}, \val_{2})$.
\end{lemma}
Intuitively, this lemma says that if $\expr_{1}$ and $\expr_{2}$ are authenticated computations, then for any proof $p$, executing $\run_{V} \expr_{1}~p$ and $\run_{I} \expr_{2}$ results in $\varphi$-related values.
When combined with our earlier results, \cref{thm:security:syntactic} follows as a corollary.

\section{Correctness}\label{sec:logrel-correctness}

Correctness of ADSs implemented against the Authentikit module type can, like security, be established using a logical relation.

\begin{theorem}[Correctness]\label{thm:correctness:syntactic}
  Let $\type$ be a primitive type.
  If
  \[
    \emptyset \mid \emptyset \proves \expr : \All \tauth, \tauthcomp : \karr{\ktype}{\ktype} . \tAuthentikitfunc~\tauth~\tauthcomp \to \tauthcomp~\type
  \]
  then for all $i \in \HashFamilyIndex$, if
  \[
    \langle \run_{P}~(\expr~\tapp~\tapp~\authentikit_{P}), (\emptyset, \emptyset, i)\rangle \cfsteps \langle (p, \val), \pstate_1 \rangle
  \]
  then there exists states $\pstate_{2}$ and $\pstate_{3}$ such that
  \begin{align*}
    &\langle \run_{V}~(\expr~\tapp~\tapp~\authentikit_{V})~p, (\emptyset, \emptyset, i)\rangle \steps \langle \Some \val, \pstate_2 \rangle \text{ and} \\
    &\langle \run_{I}~(\expr~\tapp~\tapp~\authentikit_{I}), (\emptyset, \emptyset, i)\rangle \steps \langle\val, \pstate_3\rangle
  \end{align*}
\end{theorem}
The theorem requires that $\expr$ can be syntactically typed as taking any Authentikit implementation and returning an authenticated computation.
In return, if we instantiate and run $\expr$ with the prover implementation $\authentikit_{P}$, producing some proof $p$, then $\expr$ instantiated with the verifier implementation $\authentikit_{V}$ must accept $p$.
Moreover, both the prover and the verifier yield the same output as the ideal semantics given by instantiating and running $\expr$ with $\authentikit_{I}$.

Proving this theorem raises two additional challenges beyond what was needed for the previous security theorem.
The first less serious challenge is that this theorem relates together \emph{three} programs (the prover, verifier, and ideal), instead of just two. Thus, we need to generalize the relational logic to support reasoning about three programs at once, and also generalize the logical relations model to a ternary relation.

The second more difficult challenge is that the theorem considers executions in which the proof generated by the prover is supplied as input to the verifier.
In the relational logic we have presented so far (and all other relational program logics we are aware of), there is no natural way to prove a relational specification where the output of one program should be the input to the other.
At best, one can instead essentially prove two, separate \emph{unary} specifications about the prover and verifier by first showing that all proofs generated by the prover satisfy some property $P$, and then showing that, if the property $P$ is assumed as a precondition of the input to the verifier, the verifier will succeed and return the same result.

It might be possible to make this unary approach work, but doing so gives up many of the benefits that relational program logics bring.
It has long been known in the relational program logic literature that relational proofs are usually simplified when the two programs can be \emph{aligned} or \emph{synchronized}~\citep{naumann-alignment, banerjee-relational, unno-relational}, so that the proof reasons about similar steps in the two programs at the same time.
To address this second challenge while still retaining aligned, relational reasoning, we develop a novel use of a proof technique called \emph{prophecy variables}~\citep{abadi-lamport-prophecies}.

Before proceeding to the proof, we note that the formal statement of correctness for $\lambdaAuth$ given by \citet{miller-ads} is slightly different from \cref{thm:correctness:syntactic}. In particular, their theorem says that if the ideal execution returns a value $v$, then there is an execution of the prover returning $v$ and a proof $p$ that the verifier will accept, resulting in the same value $v$.
Besides taking the execution of the ideal as a premise instead of an execution of the prover, their version of the theorem considers normal executions of the prover, whereas our version only considers collision-free traces.
However, while the formulation of correctness given by \citeauthor{miller-ads} holds for the $\lambdaAuth$ core calculus, it \emph{does not} hold when applying some of the optimizations implemented in the $\lambdaAuth$ compiler.
In contrast, as we will see in \cref{sec:optimizations}, when those corresponding optimizations are applied to an implementation of Authentikit, the formulation given in \cref{thm:correctness:syntactic} does hold. 

\subsection{Logical Relation for Correctness}
Our ternary logical relation $\Theta \mid \Gamma \vDash \expr_{1} \preceq \expr_{2} \preceq \expr_{3} : \type $ is defined using a ternary variant of the collision-free logic that encodes the verifier and ideal program as two separation logic resources $\spec_{V}(\expr)$ and $\spec_{I}(\expr)$.
These resources can be updated and progressed according to the operational semantics just as for the $\spec(\expr)$ resource introduced in \cref{sec:logic}.
Equipped with the ternary logic, we define a logical relation similar to the binary relation from \cref{sec:sec:ref} by generalization to the ternary case in the obvious way.
For instance, the term interpretation looks as follows.
\begin{align*}
  &\SemE{\relvar}(\expr_{1}, \expr_{2}, \expr_3) \eqdef{} \All i, \lctx_{2}, \lctx_{3} . \\
  & \qquad
    \begin{aligned}[t]
      &\hashindex(i) \sep \spec_{V}(\fillctx \lctx_{2} [\expr_{2}]) \sep \spec_{I}(\fillctx \lctx_{3} [\expr_{3}]) \wand{} \\
      &\wpre{\expr_{1}}
        {\val_{1} \ldotp \Exists \val_{2}, \val_{3} .
        \begin{aligned}
          &\spec_{V}(\fillctx\lctx_{2}[\val_{2}]) \sep \spec_{I}(\fillctx\lctx_{3}[\val_{3}]) \sep{} \\
          &\relvar(\val_{1}, \val_{2}, \val_{3})
        \end{aligned}
        }
    \end{aligned}
\end{align*}
The fundamental theorem follows in a similar way.
\begin{theorem}\label{thm:ternary-fundamental}
  If $\Theta \mid \Gamma \vdash \expr : \type $
  then $\Theta \mid \Gamma \vDash \expr \preceq \expr \preceq \expr : \type $.
\end{theorem}

\subsection{Correctness of Authentikit}
The correctness proof follows the same recipe as the security proof. We start by defining a suitable interpretation of $\tauth$ and $\tauthcomp$, and then show that $\authentikit_{P}$, $\authentikit_{V}$, and $\authentikit_{I}$ are related according to the ternary logical relation.
The key difference lies in the interpretation of $\tauthcomp$, the type variable for authenticated computations.
This interpretation uses prophecy variables to ``predict'' what the proof generated by the prover in the future will be

A prophecy variable is a ghost variable that supplies information about what will happen later on in execution.
In the Iris realization \cite{iris-prophecies}, a prophecy variable manifests as a separation logic resource $\isproph(\alpha, p)$, where $\alpha$ is a variable identifier and $p$ is a sequence of values.
The assertion tells us that we own the exclusive right to resolve (\ie, assign values to) $\alpha$, and that the predicated values that it will be resolved to are given by the sequence $p$.

Prophecy variables are resolved using a (ghost) programming language construct $\Resolve \alpha~s$ using the rule below.
\begin{align*}
  \inferrule[{\TirNameStyle{wp-proph-resolve}}]
  {}
  {  \isproph(\alpha, p) \vdash \wpre{\Resolve \alpha~s}{\_ \ldotp \Exists p' . p = s :: p' \sep \isproph(\alpha, p') }}
  {\label{wp-proph-resolve}}
\end{align*}
By resolving $\alpha$ to $s$, the rule reveals that $p = s :: p'$, \ie{} we may reason \emph{as if} $s$ was equal to the head of the prophecy sequence $p$, even though the value $s$ may have been the result of a computation and not statically known beforehand.
The prophecy variable, however, allows us to \emph{refer} to this future resolved value earlier on without knowing the exact value yet.

In order to prophesy the prover's output, we add a $\Resolve$ command for a designated prophecy variable $\alpha$ in the $\authentikit_{P}$ implementation of the $\unauth$ procedure.
\newcommand{\evivar}{\textit{evi}}
\begin{align*}
  \unauth_{P} \eqdef{} \lambda \evivar, (a, \_), () \ldotp
  \Let s = \evivar~a in \Resolve \alpha~s; ([s], a)
\end{align*}
Thus, the prophecy sequence $p$ for $\alpha$ will correspond to the list of values that make up the proof that is produced. 
We define $\isprophproof(\alpha, \val, p) \eqdef{} \isproph(\alpha, p) \sep \isproof(\val, p)$, as a representation predicate for this prophesied proof stream.

Then, the ternary interpretation of authenticated computations looks as follows.
\begin{align*}
  &\Sem{\tauthcomp}(\relvar)(\val_1, \val_2, \val_3) \eqdef{}
    \All \lctx_2, \lctx_3, \alpha, p, \valB . \\
  &\quad
    {
    \spec_{V}(\fillctx \lctx_{2} [\val_2~\valB]) \sep \spec_{I}(\fillctx \lctx_{3} [\val_3~\TT]) \sep \isprophproof(\alpha, \valB, p)
    } \wand \\
  &\quad \wpre
    {\val_{1} ()}
    {
    \begin{aligned}
      u \ldotp
      &\Exists p_{1}, p_{2}, \valB_{1}, \valB_{2}, a_{1}, a_{2}, a_{3} . \\
      &u = (\valB_1, a_{1}) \sep p = p_{1} \mdoubleplus p_{2} \sep \relvar(a_{1}, a_{2}, a_{3}) \sep{} \\
      &\isproof(\valB_{1},  p_{1}) \sep \isprophproof(\alpha, \valB_{2}, p_{2}) \sep{} \\
      &\spec_{V}(\fillctx \lctx_{2} [\Some~(\valB_{2}, a_{2})]) \sep \spec_{I}(\fillctx \lctx_{3} [a_{3}])
    \end{aligned}
    }
\end{align*}
In the above, the input $\valB$ to the verifier corresponds to the (prophesied) proof stream $p$.
In the postcondition of $v_1$, it is revealed that $p$ is the concatenation of two proof streams $p_{1}$ and $p_{2}$ where $p_{1}$ corresponds to the value $w_{1}$ produced by the prover during execution of $v_1$, and $p_{2}$ corresponds to the value $w_{2}$ returned by the verifier, which represents the remaining proof stream that the verifier has not yet consumed.

The definition satisfies the correctness statement shown below.
\Cref{thm:correctness:syntactic} then follows by the same procedure as for \cref{thm:security:syntactic}.
\begin{lemma}[Correctness, semantic]\label{thm:correctness:sem}
  Let $\varphi$ be a first-order ternary relation. If
  $\SemE{\Sem{\tauthcomp}~\varphi}(\expr_{1}, \expr_{2}, \expr_{3})$ then for all $i \in \HashFamilyIndex$,
  if
  \[\langle \run_{P} \expr_{1}, (\emptyset, \emptyset, i)\rangle \cfsteps \langle (p, \val_{1}), \pstate_1 \rangle\]
  then
  $\langle \run_{V} \expr_{2}~p, (\emptyset, \emptyset, i)\rangle \cfsteps \langle \Some \val_{2}, \pstate_2 \rangle$ and
  $\langle \run_{I} \expr_{3}, (\emptyset, \emptyset, i)\rangle\allowbreak \steps \langle\val_{3}, \pstate_3\rangle$ such that
  $\varphi(\val_{1}, \val_{2}, \val_{3})$.
\end{lemma}

%%% Local Variables:
%%% mode: latex
%%% TeX-master: "ccs25-auth-full-version"
%%% End:

%% file: optimizations.tex
\section{Optimizations}
\label{sec:optimizations}

This section discusses four optimizations to the Authentikit library that reduce the size of the proof stream and/or speed up proving or verification.
We show using our logical relation that each of them are secure and correct by adapting the proofs from \cref{sec:logrel-sec} and \cref{sec:logrel-correctness}.

\paragraph{Proof Accumulator}
The prover implementation from \cref{fig:authentikit-impl} uses a functional list append in the bind function to combine the proof streams produced by successive authenticated computations.
Functional list append is linear in the length of its first argument, and so repeatedly appending to the end of a list in this way leads to quadratic running time.
A standard functional programming optimization to avoid this quadratic behavior is to instead \emph{prepend} new proof items to an accumulator that gets reversed before the final proof stream is emitted.
An excerpt of the optimization is shown in \cref{fig:prover:rev}.
The verifier remains unchanged and it is therefore only the correctness proof and in turn the interpretation of authenticated computations that needs to be adapted to take the accumulator and list reversal into account.

\begin{figure}
  \centering
\begin{lstlisting}[language=ocaml]
module Prover =
  type 'a m = proof -> proof * 'a
  (* ... *)
  let unauth evi (a, _) pf = (evi a :: pf, a)
  let run m =
    let pf, res = m [] in (List.rev pf, res)
end
\end{lstlisting}
  \caption{Excerpt of the proof accumulator optimization.}
  \label{fig:prover:rev}
\end{figure}

\paragraph{Reuse Buffering}
The $\lambdaAuth$ compiler implements an optimization to reduce the size of the proof stream in cases where the same items may appear in the proof stream multiple times.
For example, a client may perform a batch of operations that end up re-traversing many of the same nodes of the ADS, in which case the hashes of these nodes do not need to be repeated multiple times.

We implement a similar optimization by modifying the prover and verifier $\unauth$ procedures.
An excerpt of the verifier modification is found in \cref{fig:verifier:buffer}.
Both the prover and the verifier maintains a cache of previously seen hashes.
When the $\unauth$ function is invoked on an authenticated value, both parties will first consult the cache.
If a hash is found, the prover omits adding the proof to the proof stream (because it is not needed), and the verifier returns the deserialization of the value found in the cache.
If the hash is not found, a proof will be emitted/consumed and the verifier deserializes and checks the proof.
If the check is successful, the result is added to the cache.

By adding proof reuse, correctness of the prover now relies on the collision-free property of the hash function: the prover may return a wrong result if a hash collides with a previously seen hash found in the cache.
As such, the correctness theorem only applies to collision-free prover execution traces.
While \citet{miller-ads} implement the optimization in the $\lambdaAuth$ compiler, their formal model only considers the unoptimized version.
In particular, their statement of the correctness theorem does not appear to hold for this optimization, as we alluded to in \cref{sec:logrel-correctness}.

To account for this optimization, our security and correctness proofs need to be adapted but the changes are minimal.
The main effort involves verifying the cache data structures and adapting the interpretation of authenticated computations to account for the cache and its contents.
For example, the verifier cache maps hashes $\hash_{i}(s)$ to $s$ where $s$ is the serialization of some authenticated value.
We record that $s$ was hashed during execution using the $\hashed(i, s)$ resource to account for hash collisions using the collision-free logic.

\begin{figure}
  \centering
\begin{lstlisting}[language=ocaml]
module Verifier =
  type 'a m = pfstate ->
              [`Ok of pfstate * 'a | `ProofFailure]
  (* ... *)
  let unauth evi h pf =
    match Map.find_opt h pf.cache with
    | None ->
       match pf.pf_stream with
       | [] -> `ProofFailure
       | p :: ps when hash p = h ->
          match evi.deserialize p with
          | None -> `ProofFailure
          | Some a ->
            `Ok ({pf_stream = ps;
                  cache = Map.add h p pf.cache}, a)
       | _ -> `ProofFailure
    | Some p ->
       match evi.deserialize p with
       | None -> `ProofFailure
       | Some a -> `Ok (pf, a)
  (* ... *)
end
\end{lstlisting}
  \caption{Excerpt of verifier version of proof-reuse buffering.}
  \label{fig:verifier:buffer}
\end{figure}

\paragraph{Heterogeneous Reuse Buffering}
The verifier implementation for the previous optimization in \cref{fig:verifier:buffer} performs deserialization when inserting an item into the cache but also when it is retrieved.
The second deserialization can be avoided by caching the authenticated value itself rather than its serialization.
However, this requires a heterogeneous cache since different authenticated values have different types.
In OCaml, this would not be syntactically well typed and in fact \citet{miller-ads} does not consider this optimization in their implementation of \lambdaAuth.
To implement the optimization in OCaml, we have to resort to the \texttt{Obj.magic} featue of OCaml that bypasses the typechecker for the cache operations.
However, even though this heterogeneous cache is not syntactically well-typed, we prove that it is safe and show that $\authentikit_{V}$ is still related to the prover and ideal versions in the logical relations at the appropriate type interface.
The optimization only requires minimal changes to the security and correctness proof.

\paragraph{Stateful Buffering}
The caching mechanisms considered in the previous sections are implemented using purely functional data structures.
For instance, the cache in \cref{fig:verifier:buffer} is passed as an argument to the $\unauth$ function.
Both OCaml and \thelang{} are expressive enough to implement and use heap-allocated data structures and so we have also implemented and verified the aforementioned optimizations using a heap-allocated cache.

%%% Local Variables:
%%% mode: latex
%%% TeX-master: "ccs25-auth-full-version"
%%% End:

%% file: semantic-proofs.tex
\section{Manual Security and Correctness Proofs}
\label{sec:semantic-proofs}

\cref{thm:security:sem} and \cref{thm:correctness:sem} hold for \emph{any} terms $\expr_{1}$, $\expr_{2}$, and $\expr_{3}$ that satisfy our interpretations of authenticated computations, \ie{}, terms for which we can derive the associated relational weakest preconditions.
These relational weakest preconditions hold automatically for all well-typed clients of the Authentikit library by the fundamental lemmas for the logical relations, which is how we deduced \cref{thm:security:syntactic} and \cref{thm:correctness:syntactic}.
However, alternatively, if we have manually-written prover and verifier implementations of an operation on an authenticated data structure that does \emph{not} use the Authentikit library, we can instead directly prove that the implementation inhabits the logical relation and therefore satisfies the relational weakest precondition.
As the logical relation is \emph{compatible} (\ie{}, satisfies the compatibility lemmas), this also means that such implementations can be soundly linked with automatically generated code for other operations on the same data structure.

We apply this methodology to an optimized \texttt{retrieve} operation on Merkle trees.
The optimization addresses a redundancy in the proof objects that \texttt{retrieve} generates, which also occurs in the corresponding $\lambdaAuth$ implementation, as noted by \citet{miller-ads}.\footnote{\citet{miller-ads} describe a compiler optimization to automatically eliminate such redundancies for $\lambdaAuth$, but the optimization does not have a correctness proof.}
For example, calling \texttt{retrieve} on the tree from \cref{fig:merkle-tree-prover} with the path $[\textsf{R}, \textsf{L}]$ produces the proof $[(h_{1}, h_{2}), (h_{5}, h_{6}), s_{5}]$ but $h_{2}$ and $h_{5}$ can be derived from the other proof items and are therefore unnecessary---the minimal proof generated by a standard, manual implementation of Merkle trees only contains $h_{1}$, $h_{6}$ and $s_{5}$.
To achieve such minimal proofs, an implementation of retrieve has to break the syntactic typing abstractions of the Authentikit library, and so we cannot automatically apply \cref{thm:security:syntactic} and \cref{thm:correctness:syntactic} to it.
Instead, we prove that such an implementation directly inhabits the logical relation at the appropriate type, \ie{}, for security we show:
\[
  \Sem{\tpath \to \ttreeauth \to \tauthcomp~(\toption~\tstring)}_{\Delta''}(\retrieve'_{V}, \retrieve_{I})
\]
where $\Delta''$ maps $\tauth$, $\tauthcomp$, and $\tpath$ to their respective interpretations; $\retrieve'_{V}$ is the optimized verifier implementation of the retrieve operation; and $\retrieve_{I}$ is an implementation of ordinary binary tree retrieval.
We show a similar statement to establish correctness.
Both the security and correctness proofs follow by induction on the path and rely on finding a suitable inductive invariant.
We refer to \cref{app:merkle-retrieve} for a description of the optimized implementation and the Coq formalization for details on the proof.

%%% Local Variables:
%%% mode: latex
%%% TeX-master: "ccs25-auth-full-version"
%%% End:

%% file: work.tex
\section{Related Work}
\label{sec:work}

\paragraph{Verification of Authenticated Data Structures.}

Merkle trees are the canonical ADS but \citet{miller-ads} also use \lambdaAuth to implement Red-black+ trees, skip lists, and planar-separator trees, among others.
All of these data structures are directly portable to Authentikit.
\citet{ads-append-only-skip-list} verify a particular notion of authenticated append-only skip lists in Agda and indicate that it does not seem possible to encode their implementation in \lambdaAuth because of the type-directed hashing discipline.
It would be interesting future work to apply our semantic approach to show that their implementation can be safely linked with code that is generated using Authentikit.
\citet{ads-formally} mechanize the proofs of security and correctness for the core calculus of $\lambdaAuth$ given by \citeauthor{miller-ads} in Isabelle/HOL.
They identify and resolve several minor technical issues in the original proofs.
As discussed in \cref{sec:introduction}, the \lambdaAuth approach---and therefore also \citeauthor{ads-formally}'s formalization ---has three important limitations as compared to our work: (1) a custom compiler frontend is needed, (2) optimizations are not covered by the security/correctness theorems, and (3) hand-written optimizations cannot be verified and integrated with automatically generated code.

\citet{merkle-functors} show how to systematically and modularly derive ADSs as data types in Isabelle/HOL using so-called Merkle functors.
The construction comes with no formal security or correctness guarantees. They point out that HOL's lack of abstraction over type constructors (which are supported by OCaml and \thelang) hinders expressing their process in its full generality.

\citet{merkle-tree-f-star} implement and verify a variant of Merkle Patricia Trees in F*.
They show that each of their tree-manipulating functions are functionally correct and that the hashing scheme is collision resistant by a reduction to collision resistance of the hash function being used.
\citet{f-star-fastver2} use a sparse, incremental Merkle tree to authenticate the state of a verified key-value store and show sequential consistency up to hash collision.
In contrast to both of these efforts, our proof guarantees security and correctness for \emph{all} syntactically well-typed functions.

\paragraph{Logical Relations}

Relational parametricity and free theorems for languages with higher kinds have been studied using several different approaches~\citep{hasegawa-relational-parametricity, reflexive-graphs-parametricity, atkey-relational-parametricity, weirich-vytiniotis-parametricity, voigtlander-free-theorems}.
Our logical-relations model takes inspiration from recent work~\citep{sieczkowski-gadt} that developed a model of a type system with generalized algebraic data types.
In addition, our Coq formalization builds on their formalization of intrinsically well-kinded types that uses a notion of functorial syntax~\citep{functorial-syntax}.

Logical relations have been used to establish a variety of different security properties of expressive type systems.
\citet{seloc}~develop a relation to establish a timing-sensitive notion of noninterference for an information-flow control type system.
\citet{iris-tini}~similarly develop a model for a termination-insensitive notion of noninterference.
\citet{cerise-oopsla, cerise-jacm, strydonck-capability} and \citet{swasey-capability}~use logical relations to establish so-called robust safety properties that formalize the security guarantee offered by a capability machine.
\citet{mswasmcert}~use a logical relation to show robust capability safety of a WebAssembly extension and \citet{sandboxing-sammler}~consider a sandboxing mechanism.

\paragraph{Prophecy Variables}

Prophecy variables were originally developed by~\citet{abadi-lamport-prophecies} to establish certain program refinements that require speculative reasoning about future events. They are commonly used in concurrent program verification for reasoning about future-dependent linearization points but have since then also found use in relational reasoning that requires alignment \cite{naumann-alignment, unno-relational}.
\citet{reloc}~show how to integrate prophecy variables in a logical relation for contextual refinement of concurrent programs.
\citet{vilhena-spy-game} consider several (unary) examples where prophecy variables are seemingly required in verifying deterministic and sequential code, including a proof of a structural infinitary conjunction rule for separation logic.
To our knowledge, our work is the first to use prophecy variables to show free theorems.

%%% Local Variables:
%%% mode: latex
%%% TeX-master: "ccs25-auth-full-version"
%%% End:

%% file: conclusion.tex
\section{Conclusion}
\label{sec:conclusion}

Authenticated data structures allow untrusted third parties to carry out operations which produce proofs that can be used to check that the results of the operation are valid.
In this work, we showed how the Authentikit library generates secure and correct authenticated data structures automatically.

Our proof uses a new relational separation logic for reasoning about programs that use collision-resistant cryptographic hash functions.
We use the logic as a basis for constructing two logical-relations models of \thelang{} that are expressive enough to show security and correctness of authenticated data structures generated using Authentikit.
The correctness proof, in particular, relies on a novel use of prophecy variables.
Finally, we also showed how to use our models to prove security and correctness of four optimizations to the library and how optimized, hand-written implementations of authenticated data structures can be soundly linked with automatically generated code.

%%% Local Variables:
%%% mode: latex
%%% TeX-master: "ccs25-auth-full-version"
%%% End:

%% file: appendix.tex
\section[Syntax and Type System of F-omega-mu-ref]{Syntax and Type System of \thelang}\label{app:language}

\cref{fig:language} shows the full syntax and an excerpt of the typing rules of \thelang.
Types have a standard kind structure: the kind $\ktype$ for the kind of proper types and $\karr{\kind_{1}}{\kind_{2}}$ for constructors that given a type of kind $\kind_{1}$ produce a type of kind $\kind_{2}$.
Well-kinded types are formed according to the judgment $\Theta \vdash \type : \kind$ where $\Theta$ assigns kinds to type variables.
The typing judgment $\Theta \mid \Gamma \vdash \expr : \type$ assigns a type of kind $\ktype$ to a term in contexts $\Theta$ and $\Gamma$.
The context $\Theta$ assigns kinds to type variables and $\Gamma$ assigns types of kind $\ktype$ to term-level variables.
Note that we use \emph{type elimination contexts} \cite{rossberg-iso} to succinctly express well-kinded use of (higher-kinded) iso-recursive types.
A well-kinded type elimination context $\Theta \vdash \tctx : \kind_1 \hookrightarrow \kind_2 $ takes as input a type of kind $\kind_{1}$ and produces a type of kind $\kind_{2}$.
The type equivalence relation $\Theta \vdash \type \equiv \typeB : \kind$ permits one to replace the type of a term with any type equivalent to it.

\begin{figure*}[p]
  \centering
\textbf{Syntax}
  \begin{alignat*}{4}
    &  \text{(values)} \hspace{3em}
    && \val \hspace{1em}
    && \bnfdef{}\hspace{1em}
    && \TT \ALT
       b \in \bool \ALT
       z \in \integer \ALT
       s \in \String \ALT
       \loc \in \Loc \ALT
       (\val,\val) \ALT
       \Inl \val  \ALT
       \Inr \val \ALT \Rec f \var = \expr \ALT
       \Lambda \expr \ALT
       \Pack \val
    \\
    &  \text{(terms)}
    && \expr
    && \bnfdef{}
    && \val \ALT
       \var \in \Var \ALT
       \HLOp_{1} \expr \ALT
       \expr \HLOp_{2} \expr \ALT
       \expr~\expr \ALT
       \If \expr then \expr \Else \expr \ALT
       \Fst \expr \ALT
       \Snd \expr \ALT
       \Case~\expr~\expr~\expr \ALT
       % \Match \expr with \Inl \val~ => \expr | \Inr \valB => \expr end \ALT
       \Alloc \expr \ALT
       \deref \expr \ALT
       \expr \gets \expr \ALT
    \\ & && &&
    && \fold \expr \ALT
       \unfold \expr \ALT
       \expr~\tapp \ALT
       \Pack \expr \ALT
       \Unpack \expr as \var in \expr \ALT
       \Hash \expr
    \\
    &  \text{(unary operators)}\hspace{1em}
    && \HLOp_{1}
    && \bnfdef{}
    && - \ALT ! \ALT \textlang{intOfString} \ALT \textlang{stringOfInt} \ALT \textlang{length}
    \\
    &  \text{(binary operators)}\hspace{1em}
    && \HLOp_{2}
    && \bnfdef{}
    && + \ALT - \ALT \cdot \ALT = \ALT \mdoubleplus \ALT \ldots
    \\
    &  \text{(evaluation contexts)}\hspace{1.5em}
    && \lctx
    && \bnfdef{}
    && - \ALT
       \HLOp_{1} \lctx \ALT
       \expr \HLOp_{2} \lctx \ALT
       \lctx \HLOp_{2} \val \ALT
       \expr~\lctx \ALT
       \lctx~\val \ALT
       \Alloc \lctx \ALT
       \deref \lctx \ALT
       \expr \gets \lctx \ALT
       \lctx \gets \val \ALT
       \ldots
    \\
    &  \text{(kinds)}
    && \kind
    && \bnfdef{}
    && \ktype \ALT
       \karr{\kind}{\kind}
    \\
    &  \text{(types)}
    && \type
    && \bnfdef{}
    && \tvar \in \TyVar \ALT
       \lambda \tvar : \kind \ldotp \type \ALT
       \type~\type \ALT
       \tconstr
    \\
    & \text{(constructors)}
    && \tconstr
    && \bnfdef{}
    && \tunit \ALT
       \tbool \ALT
       \tint \ALT
       \tstring \ALT
       \times \ALT
       + \ALT
       \to \ALT
       \tref  \ALT
       \forall_{\kind} \ALT
       \exists_{\kind} \ALT
       \mu_{\kind}
    \\
    & \text{(type elim. contexts)}\hspace{1em}
    && \tctx
    && \bnfdef{}
    && - \ALT T~\type
  \end{alignat*}
  \textbf{Type Formation}
  \begin{mathpar}
    \infer{ }
    {\tunit, \tbool, \tint, \tstring ~:~ \ktype }
    \and
    \infer{ }
    {\times, +, \to ~:~ \karr{\ktype}{\karr{\ktype}{\ktype}}}
    \and
    \infer{ }
    {\tref ~:~ \karr{\ktype}{\ktype}}
    \and
    \infer{ }
    {\forall_{\kind}, \exists_{\kind} ~:~ \karr{(\karr{\kind}{\ktype})}{\ktype}}
    \and
    \infer{ }
    {\mu_{\kind} ~:~ \karr{(\karr{\kind}{\kind})}{\kind}}
  \end{mathpar}
  \begin{mathpar}
    \infer
    { \tvar : \kind \in \Theta }
    {\Theta \vdash \tvar : \kind}
    \and
    \infer
    { \tconstr : \kind }
    { \Theta \vdash \tconstr : \kind }
    \and
    \infer
    { \Theta, \tvar : \kind_1 \vdash \type : \kind_2 }
    { \Theta \vdash \lambda \tvar : \kind_1 . \type : \karr{\kind_1}{\kind_2} }
    \and
    \infer
    { \Theta \vdash \typeB : \karr{\kind_1}{\kind_2} \\
      \Theta \vdash \type  : \kind_1 }
    { \Theta \vdash \typeB~\type : \kind_2 }
  \end{mathpar}
  \textbf{Type Elimination Context Formation}
  \begin{mathpar}
    \infer
    { }
    {\Theta \vdash - : \kind \hookrightarrow \kind}
    \and
    \infer
    { \Theta \vdash \tctx : \kind_1 \hookrightarrow (\karr{\kind}{\kind_2}) \\
      \Theta \vdash \type : \kind
    }
    { \Theta \vdash \tctx~\type : \kind_1 \hookrightarrow \kind_2 }
  \end{mathpar}
  \textbf{Type Equivalence}
  \begin{mathpar}
    \infer
    { \Theta \vdash \type : \kind }
    { \Theta \vdash \type \equiv \type : \kind }
    \and
    \infer
    { \Theta \vdash \type' \equiv \type : \kind
      % \\ \Theta \vdash \type, \type' : \kind
    }
    { \Theta \vdash \type \equiv \type' : \kind }
    \and
    \infer
    { \Theta \vdash \type \equiv \type' : \kind \\
      \Theta \vdash \type' \equiv \type'' : \kind
      % \Theta \vdash \type, \type', \type'' : \kind
    }
    { \Theta \vdash \type \equiv \type'' : \kind }
    \and
    \infer
    { \Theta, \tvar : \kind_1 \vdash \type \equiv \type' : \kind_2
      % \Theta, \tvar : \kind_1 \vdash \type, \type' : \kind_2
    }
    { \Theta \vdash \lambda \tvar : \kind_1 \ldotp \type \equiv \lambda \tvar : \kind_1 \ldotp \type' : \karr{\kind_1}{\kind_2} }
    \and
    \infer
    { \Theta \vdash \type \equiv \typeB : \karr{\kind_1}{\kind_2} \\
      \Theta \vdash \type' \equiv \typeB' : \kind_1
      % \Theta \vdash \type, \typeB : \karr{\kind_1}{\kind_2} \\
      % \Theta \vdash \type', \typeB' : \kind_1
    }
    { \Theta \vdash \type~\type' \equiv \typeB~\typeB' : \kind_2}
    \and
    \infer
    { \Theta, \tvar : \kind_1 \vdash \type : \kind_2 \\
      \Theta \vdash \typeB : \kind_1 }
    { \Theta \vdash (\lambda \tvar : \kind_1 \ldotp \type) \typeB \equiv \subst{\typeB}{\tvar}{\type} : \kind_2}
    \and
    \infer
    { \Theta \vdash \type : \karr{\kind_1}{\kind_2} \\ \tvar \not\in \Theta }
    { \Theta \vdash \type \equiv \lambda \tvar : \kind_1 \ldotp \type~\tvar : \karr{\kind_1}{\kind_2} }
  \end{mathpar}
  \textbf{Term Formation (excerpt)}
  \begin{mathpar}
    \infer
    { \Gamma(x) = \type }
    { \Theta \mid \Gamma \vdash \var : \type }
    % \and
    % \infer
    % { \Theta \mid \Gamma \vdash \expr_1 : \type_1 \\
    %   \Theta \mid \Gamma \vdash \expr_2 : \type_2 \\
    %   \overline{\HLOp_2} = \type_1 \to \type_2 \to \type }
    % { \Theta \mid \Gamma \vdash \expr_1 \HLOp_2 \expr_2 : \type }
    \and
    \infer
    {\Theta \mid \Gamma, f : \type_1 \to \type_2, \var : \type_1 \vdash \expr : \type_2 \\
      \Theta \vdash \type_1, \type_2 : \ktype }
    { \Theta \mid \Gamma \vdash \Rec f \var = \expr : \type_1 \to \type_2 }
    \and
    \infer
    { \Theta \mid \Gamma \vdash \expr_2 : \type_1 \to \type_2 \\
      \Theta \mid \Gamma \vdash \expr_1 : \type_1 \\
    }
    { \Theta \mid \Gamma \vdash \expr_2~\expr_1 : \type_2 }
    \\
    \and
    \infer
    { \Theta, \tvar : \kind \mid \Gamma \vdash \expr : \type }
    { \Theta \mid \Gamma \vdash \Lambda \expr : \All \tvar : \kind . \type }
    \and
    \infer
    { \Theta \mid \Gamma \vdash \expr : \All \tvar : \kind . \type \\
      \Theta, \tvar : \kind \vdash \type : \ktype \\
      \Theta \vdash \typeB : \kind
    }
    { \Theta \mid \Gamma \vdash \expr~\tapp : \subst{\type}{\tvar}{\typeB} }
    \and
    \infer
    { \Theta \mid \Gamma \vdash \expr : \subst{\type}{\tvar}{\typeB} \\
      \Theta, \tvar : \kind \vdash \type : \ktype \\
      \Theta \vdash \typeB : \kind
    }
    { \Theta \mid \Gamma \vdash \Pack \expr : \Exists \tvar : \kind . \type }
    \and
    \infer
    { \Theta \mid \Gamma \vdash \expr_1 : \Exists \tvar : \kind . \type_1 \\
      \Theta, \tvar : \kind \mid \Gamma, \var : \type_1 \vdash \expr_2 : \type_2 \\
      \Theta, \tvar : \kind \vdash \type_1 : \ktype \\
      \Theta \vdash \type_2 : \ktype \\
      \tvar \not\in \Theta, \type_2
    }
    { \Theta \mid \Gamma \vdash \Unpack \expr_1 as \var in \expr_2 : \type_2 }
    \\
    \and
    \infer
    { \Theta \mid \Gamma \vdash \expr : \fillctx \tctx [\subst{\type}{\tvar}{\mu \tvar : \kind \ldotp \type}] \\
      \Theta \vdash \tctx : \kind \hookrightarrow \ktype \\
      \Theta, \tvar : \kind \vdash \type : \kind }
    { \Theta \mid \Gamma \vdash \fold \expr : \fillctx \tctx [\mu \tvar : \kind \ldotp \type] }
    \and
    \infer
    { \Theta \mid \Gamma \vdash \expr : \fillctx \tctx [\mu \tvar : \kind \ldotp \type] \\
      \Theta \vdash \tctx : \kind \hookrightarrow \ktype \\
      \Theta, \tvar : \kind \vdash \type : \kind }
    { \Theta \mid \Gamma \vdash \unfold \expr : \fillctx \tctx [\subst{\type}{\tvar}{\mu \tvar : \kind \ldotp \type}] }
    \and
    \infer
    { \Theta \mid \Gamma : \expr : \type }
    { \Theta \mid \Gamma \vdash \Alloc \expr : \tref\type }
    \and
    \infer
    { \Theta \mid \Gamma : \expr_1 : \tref\type \\
      \Theta \mid \Gamma : \expr_2 : \type }
    { \Theta \mid \Gamma \vdash \expr_1 \gets \expr_2 : \tunit }
    \and
    \infer
    { \Theta \mid \Gamma : \expr : \tref\type }
    { \Theta \mid \Gamma \vdash \deref \expr : \type }
    \and
    \infer
    { \Theta \mid \Gamma \vdash \expr : \tstring }
    { \Theta \mid \Gamma \vdash \Hash \expr : \tstring }
    \and
    \infer
    { \Theta \mid \Gamma \vdash \type \equiv \typeB : \ktype \\
      \Theta \mid \Gamma \vdash \expr : \typeB }
    { \Theta \mid \Gamma \vdash \expr : \type }
  \end{mathpar}
  \caption{Syntax and type system of \thelang.}
  \label{fig:language}
\end{figure*}

\section{Optimized Retrieve Operation}\label{app:merkle-retrieve}

Recall from \cref{sec:semantic-proofs} that the proofs generated by \texttt{retrieve} have some redundancy and are larger than necessary.
For example, calling \texttt{retrieve} on the tree from \cref{fig:merkle-tree-prover} with the path $[\textsf{R}, \textsf{L}]$ produces the proof $[(h_{1}, h_{2}), (h_{5}, h_{6}), s_{5}]$, but $h_{2}$ and $h_{5}$ can be computed by hashing the other elements of the proof stream.
A standard, manual implementation of Merkle trees therefore produces a proof that only contains $h_{1}$, $h_{6}$ and $s_{5}$.
These are essentially the hashes of the siblings of the nodes traversed by the path in the tree, followed by an unhashed version of the last-visited tree node.
In this section, we describe our optimized implementation of the Merkle tree retrieve operation, which we call \texttt{retrieve'} henceforth for brevity, which produces and consumes these minimal proofs.

Since we implement this optimized functionality directly instead of using the generic \texttt{Authentikit} interface, we have two separate implementations for the prover and verifier.
These implementations are organized into modules with the signature \texttt{MERKLE\_RETRIEVE}.
Although we do not use the \texttt{auth} and \texttt{unauth} operations from \texttt{Authentikit} in the optimized retrieve, we still parameterize these module by an implementation of \texttt{Authentikit} so that they serialize and manipulate proof objects in a way that will be compatible with code written using \texttt{Authentikit}.

\cref{fig:prover:merkle:magic} gives the implementation of the prover.
The function \texttt{retrieve\_prover\_aux} is the core method, which descends through the tree and constructs the proof for the retrieval.
Because we want to use our optimized operations alongside operations implemented using \texttt{Authentikit}, we need a way to combine the proof that \texttt{retrieve\_prover\_aux} returns with the larger proof stream being produced by the \texttt{Authentikit} operations.
To do so, \texttt{retrieve'} uses a function called \texttt{push\_proof} to insert the proof returned by \texttt{retrieve\_prover\_aux}.

The verifier's version of the \texttt{retrieve'} function is shown in the \cref{fig:verifier:merkle:magic}.
It uses a function \texttt{pop\_proof} to retrieve the next proof from the proof stream.
Recall that the prover adds the unhased version of the node at the end of the path.
The verifier hashes this, and then using the rest of the proof, it recursively calls \texttt{retrieve\_verifier\_aux} to builds the hashes of all the sub-trees along the path, eventually recomputing the root hash of the tree.
If the computed hash of the entire tree, matches the hash the verifier is already provided with, the verifier accepts the proof.

Finally, we show in \cref{fig:merkle:modification:magic}, how we modify the \texttt{Merkle} functor from \cref{fig:merkle} so that it now accepts another argument of type \texttt{MERKLE\_RETRIEVE}, which provides the hand-written \texttt{retrieve'} function.
When instantiating the \texttt{Merkle} functor with the prover or verifier implementations of Authentikit, we also instantiate it with \texttt{Merkle\_retrieve\_prover} and \texttt{Merkle\_retrieve\_verifier}, respectively.

\begin{figure*}[p]
  \centering
\begin{lstlisting}[language=ocaml]
module Merkle_retrieve_prover : MERKLE_RETRIEVE =
  functor (K: AUTHENTIKIT) -> struct
    open K

    type path = [`L|`R] list
    type tree = [`leaf of string | `node of tree auth * tree auth]
    type tree_auth = tree auth
    let tree = Authenticatable.(sum string (pair auth auth))
    let append_to_list_proof s proof = (* ... *)

    let rec retrieve_prover_aux path t r_proof =
      match path, t with
      | [], (`leaf s, h) ->
        append_to_list_proof (`leaf s) r_proof, Some s
      | _, (`leaf s, h) ->
        append_to_list_proof (`node (`leaf s)) r_proof, None
      | [], (`node ((l, h_l), (r, h_r)), h) ->
        append_to_list_proof (`node (`node (h_l, h_r))) r_proof, None
      | `L::path, (`node (l, (r, h_r)), h) ->
        append_to_list_proof h_r r_proof |> retrieve_prover_aux path l
      | `R::path, (`node ((l, h_l), r), h) ->
        append_to_list_proof h_l r_proof |> retrieve_prover_aux path r

    let retrieve' =
      Obj.magic (fun path t proof ->
        let retrieve_proof, a = retrieve_prover_aux path t empty_list_proof in
        push_proof (Obj.magic (retrieve_proof)) proof, a)
  end
\end{lstlisting}
  \caption{Manually-written retrieve code for the prover.}
  \label{fig:prover:merkle:magic}
\end{figure*}

\begin{figure*}[p]
  \centering
\begin{lstlisting}[language=ocaml,escapechar=/]
module Merkle_retrieve_verifier : MERKLE_RETRIEVE =
  functor (K: AUTHENTIKIT) -> struct
    open K

    type path = [`L|`R] list
    type tree_auth = string
    let tree = Authenticatable.(sum string (pair auth auth))
    let ver_auth = (Obj.magic auth : _ -> _ -> string)

    let deserialize_list_proof proof = (* ... *)

    let rec retrieve_verifier_aux path retrieve_proof cur_hash =
      match retrieve_proof, path with
      | [], [] -> Some cur_hash
      | [], _ -> None
      | _, [] -> None
      | proof_head::proof_tail, path_head::path_tail ->
        let node_hash = retrieve_verifier_aux path_tail proof_tail cur_hash in
        match node_hash with
        | None -> None
        | Some node_hash ->
          match path_head with
          | `L -> Some (ver_auth tree (`node (node_hash, proof_head)))
          | `R -> Some (ver_auth tree (`node (proof_head, node_hash)))

    let retrieve' =
      Obj.magic (fun path t_hash proof ->
        match pop_proof proof with
        | None -> `ProofFailure
        | Some (retrieve_proof_ser, rem_proof) ->
          let proof_list = deserialize_list_proof (Obj.magic retrieve_proof_ser) in
          match proof_list with
          | [] -> `ProofFailure
          | first_element :: proof_list ->
            let result, start_hash, path, imm_return =
              match first_element with
              | `leaf s ->
                Some s, ver_auth tree (`leaf s), path, false
              | `node (`leaf s) ->
                let trunc_path = List.take (List.length proof_list) path in
                None, ver_auth tree (`leaf s), trunc_path,
                  List.length path <= List.length trunc_path
              | `node (`node v) ->
                None, ver_auth tree (`node v), path, false
            in
            if imm_return then `ProofFailure else
            match retrieve_verifier_aux path (List.rev proof_list) start_hash with
            | None -> `ProofFailure
            | Some node_hash ->
              if String.equal node_hash t_hash then `Ok (rem_proof, result)
              else `ProofFailure )
  end
\end{lstlisting}
  \caption{Manually-written retrieve code for the verifier.}
  \label{fig:verifier:merkle:magic}
\end{figure*}

\begin{figure*}[p]
  \centering
\begin{lstlisting}[language=ocaml]
module Merkle =
  functor (K : AUTHENTIKIT) -> functor (R: MERKLE_RETRIEVE) -> sig
    open K
    open R (K)

    let retrieve' path t = (Obj.magic retrieve') path t

    (* ... *)
  end
\end{lstlisting}
  \caption{Modified Merkle tree module using the manually-written retrieve.}
  \label{fig:merkle:modification:magic}
\end{figure*}

%%% Local Variables:
%%% mode: latex
%%% TeX-master: "ccs25-auth-full-version"
%%% End: